\documentclass[11pt,preprint,unsortedaddress,amsmath,amssymb,aip,jcp]{revtex4}
\usepackage{graphicx}
\usepackage{color}
\usepackage[version=3]{mhchem}
\usepackage{dcolumn}
\usepackage{multirow}
\usepackage{bm}
\usepackage[normalem]{ulem}

\usepackage{cmbright}
\begin{document}

\newcommand{\bra}[1]{\langle #1|}
\newcommand{\bs}{\boldsymbol}
\newcommand{\ket}[1]{|#1\rangle}
\newcommand{\braket}[2]{\langle #1|#2\rangle}

\title{An Exact Factorization Perspective on Quantum Interferences in Nonadiabatic Dynamics}
\author{Basile F. E. Curchod}
\email{basile.curchod@gmail.com}
\author{Federica Agostini}
\email{agostini@mpi-halle.mpg.de}
\author{E. K. U. Gross}
\affiliation{Max-Planck-Institut f\"{u}r Mikrostrukturphysik, Weinberg 2, 06120 Halle, Germany}
\date{\today}
\begin{abstract}

Nonadiabatic quantum interferences emerge whenever nuclear wavefunctions in different electronic states meet and interact in a nonadiabatic region. In this work, we analyze how nonadiabatic quantum interferences translate in the context of the exact factorization of the molecular wavefunction. In particular, we focus our attention on the shape of the time-dependent potential energy surface -- the exact surface on which the nuclear dynamics takes place. We use a {\color{black}one-dimensional} exactly-solvable model to reproduce different conditions for quantum interferences{\color{black}, whose characteristic features already appear in one-dimension}. The time-dependent potential energy surface develops complex features when strong interferences are present, in clear contrast to the observed behavior in simple nonadiabatic crossing cases. Nevertheless, independent classical trajectories propagated on the exact time-dependent potential energy surface reasonably conserve a distribution in configuration space that mimics the one of the exact nuclear probability density.

\end{abstract}
\maketitle

\section{Introduction}
\label{intro}

The molecular time-dependent Schr\"{o}dinger equation represents a Rosetta stone for a theoretical understanding of photochemical and photophysical processes, when the Born-Oppenheimer (BO) approximation~\cite{born1927quantentheorie} breaks down. The combination of important electron-nuclear couplings and nuclear quantum effects in nonadiabatic dynamics make a radical change to the simple picture the BO approximation offers for molecules in their electronic ground state~\cite{tully2000perspective}. To approximate a solution to the time-dependent Schr\"{o}dinger equation, the total molecular wavefunction is commonly expressed in a basis of BO electronic states, leading to the concept of potential energy surfaces. Based on this picture, coupled time-dependent nuclear equations -- one for each electronic state contribution --can be solved for small molecules or for a reduced-dimensionality representation of larger systems~\cite{Beck2000,worth04,persico2014overview}. A plethora of different techniques have been developed to approximate the nonadiabatic nuclear dynamics of molecules, \color{black}based for example on classical or quantum trajectories~\cite{preston71,tully90,coker1995methods,Bittner1995,prezhdo97,Tully1998,kapral:8919,burghardt2001hydrodynamic,Wyatt2001a,marxbook,Horenko2002,doi:10.1146/annurev.physchem.57.032905.104702,Jasper2006,barbatti2011,zamstein2012non,zamstein2,Gorshkov2013,curchod2013ontrajectory,persico2014overview,albareda2014correlated,guilleandivano}, (frozen) Gaussian basis sets~\cite{martinez1996multi,Ben-Nun1998,toddaims,heller2006guided,worth2008solving,richings2015quantum,shalashilin2009quantum}, linearized nonadiabatic dynamics (LAND-map)~\cite{Bonella2005}, or semiclassical considerations~\cite{herman84,sun1997,PhysRevA.71.032511}.\color{black}

The BO picture to excited-state dynamics is nonetheless not the only possible one. The molecular wavefunction can for example be represented \textit{exactly} by a simple factorization~\cite{Gross_PRL2010,Gross_JCP2012} in terms of a time-dependent nuclear wavefunction and a time-dependent electronic wavefunction, parametrically dependent on the nuclear positions. When inserted into the molecular time-dependent Schr\"{o}dinger equation, the Exact Factorization (EF) leads to coupled equations driving the dynamics of the two components of the wavefunction: a time-dependent Schr\"{o}dinger equation~\cite{Gross_PRL2013, Gross_MP2013, Gross_JCP2015, Suzuki_PCCP2015} describes the evolution of the nuclear wavefunction, where the effect of the electrons is fully accounted for by a time-dependent vector potential and a time-dependent scalar potential (or time-dependent potential energy surface, TDPES); electronic dynamics is generated by an evolution equation where the coupling to the nuclei is expressed by the so-called electron-nuclear coupling operator~\cite{Agostini_ADP2015,Scherrer_JCP2015,Schild_JPCA2016, Agostini_arXiv2016, Scherrer_arXiv2016}. {\color{black}The EF has been developed both in the time-independent~\cite{Hunter_IJQC1974, Hunter_IJQC1975_1, Hunter_IJQC1975_2, Hunter_MP1975, Bishop_MP1975, Hunter_IJQC1980, Hunter_IJQC1981, Hunter_IJQC1982, Hunter_IJQC1986, Cederbaum_JCP2013, Gross_PTRSA2014, Min_PRL2014, Cederbaum_JCP2014, Cederbaum_CP2015, Lefebvre_JCP2015_1, Lefebvre_JCP2015_2, Requist_PRA2015} and in the time-dependent~\cite{Gross_PRL2010, Gross_JCP2012, Gross_PRL2013, Gross_MP2013, Gross_EPL2014, Gross_JCP2014, Agostini_ADP2015, Gross_JCP2015, Suzuki_PCCP2015, Gross_PRL2015, Gross_JCTC2016} versions, and analyzed under different perspectives~\cite{Sutcliffe_TCA2012, Suzuki_PRA2014, Khosravi_PRL2015, Ghosh_MP2015, Scherrer_JCP2015, Schild_JPCA2016, Agostini_arXiv2016, Scherrer_arXiv2016}.} When nuclear dynamics undergoes a single nonadiabatic event, we have pointed out the properties of the TDPES and related them to the, more standard, picture provided in the BO framework, i.e., BO nuclear wavefunctions evolving on multiple static potential energy surfaces (PESs). In this situation, the TDPES shows (i) a diabatic shape in the vicinity of an avoided crossing, smoothly connecting the BO PESs involved in the process, and (ii) dynamical steps bridging piecewise adiabatic shapes, far from the avoided crossing. In particular, the steps of the TDPES have been related to the spatial splitting of the (exact) nuclear wavefunction, which reproduces the dynamics of BO wavefunctions branching in different adiabatic states. Furthermore, we have employed these observations to investigate the suitability of the (quasi)classical treatment~\cite{Gross_MP2013,Gross_JCP2015} of nuclear dynamics in situations where the electronic effect can be taken into account exactly, with the aim of proposing trajectory-based approximation schemes~\cite{Gross_EPL2014,Gross_JCP2014,Gross_PRL2015,Gross_JCTC2016} to the quantum-mechanical problem.

In the present work, we address a new problem in the context of the EF, analyzing the appearance of quantum interferences in nonadiabatic processes and identifying their signatures in the TDPES. 
Nonadiabatic quantum interferences represent a clear Achille's heel for approximate nonadiabatic methods based on classical trajectories~\cite{tully90}, such as trajectory surface hopping (TSH)~\cite{granucci1995coherent,romstad1997nonadiabatic,donoso2000simulation}. These events are for example observed whenever multiple crossings through nonadiabatic regions take place, for which a proper description of decoherence effects is paramount (examples of decoherence corrections for TSH can be found in Refs.~\cite{Granucci2007,granucci2010including,shenvi2011phase,subotnik2011decoherence,shenvi2011simultaneous,subotnik2011new,shenvi2012achieving}). 
The aim of this study is to translate the properties usually observed in the BO framework to the language of the EF and, from this, to understand the key features to be accounted for when developing approximations to the coupled electron-nuclear quantum dynamics. The reason to focus on the TDPES is  that it represents the very (and only, when the vector potential can be gauged away) quantity driving the nuclear dynamics, and all its features are intimately related to the features of the nuclear wavefunction. In order to study quantum interferences, we construct a situation where two BO nuclear wavefunctions, associated to different adiabatic states, meet at an avoided crossing. The BO wavefunctions reach the nonadiabatic region after having evolved on different paths, thus the recombination patterns might be strongly affected by the different histories. One of the questions addressed in this work is: Can such recombination patterns, i.e., nonadiabatic quantum interferences, be identified in the TDPES? The importance of capturing the recombination process imposes that the analysis shall focus on the nonadiabatic region, by contrast to previous analysis~\cite{Gross_PRL2013,Gross_JCP2015} that has instead mainly placed the attention on the TDPES far from an avoided crossing. Furthermore, as interferences are fundamentally quantum-mechanical effects, the natural question arises: Can classical trajectories, also in this case, correctly reproduce nuclear dynamics if the exact TDPES is known? The answer to this second question will potentially shed light on the possibility of capturing quantum interferences within a quantum-classical description of the nonadiabatic process. 

Aiming at answering these questions, we organize the paper as follows. In Sec.~\ref{theory} we recall the EF and its connection to the BO framework, while in Sec.~\ref{interference} we introduce the context of our analysis and define the central quantity of our study, i.e., quantum interferences. The {\color{black}one-dimensional} system under investigation and the computational details are described in Sec.~\ref{model}. The analysis is reported in Sec.~\ref{results}: we highlight the effect of quantum interferences on the nonadiabatic process in Sec.~\ref{tdpesshape}, identifying different situations where the outcome of the dynamics is more or less affected by interferences; we pinpoint the features of the TDPES related to quantum interferences in Sec.~\ref{tdpesanalysis}; we study the performance of (independent) classical trajectories evolving on the exact TDPES in reproducing nuclear dynamics in Sec.~\ref{cltrajs}. We state our conclusions in Sec.~\ref{conclusion}.

\section{Theoretical considerations}

\subsection{Exact factorization of the molecular wavefunction}
\label{theory}

In excited-state dynamics, the central equation is the time-dependent Schr\"{o}dinger equation for a molecular system,
\begin{equation}
\hat{H}(\bs r, \bs R)\Psi(\bs r, \bs R, t) = i\hbar \partial_t \Psi(\bs r, \bs R, t)\, , 
\label{tdse}
\end{equation}
where $\Psi(\bs r, \bs R, t)$ is the time-dependent molecular wavefunction with $\bs r$ and $\bs R$ being collective variables for the electronic and nuclear coordinates, respectively. In Eq.~\eqref{tdse}, $\hat{H}(\bs r, \bs R)$ is the molecular Hamiltonian, defined as the sum of the nuclear kinetic energy operator $\hat{T}_n$ and the BO Hamiltonian $\hat{H}_{BO}(\bs r, \bs R)$, i.e.,
\begin{align}
\hat{H}(\bs r, \bs R) & = \hat{T}_n + \hat{H}_{BO}(\bs r, \bs R)  \notag \\
& = \hat{T}_n +  \hat{T}_{e} + \hat{V}_{ee}(\bs r) + \hat{V}_{en}(\bs r, \bs R) + \hat{V}_{nn}(\bs R) \, .
\label{hamiltonian}
\end{align}
$\hat{V}_{ee}(\bs r)$, $\hat{V}_{en}(\bs r, \bs R)$, and $\hat{V}_{nn}(\bs R)$ are electron-electron, electron-nucleus, and nucleus-nucleus interaction potentials.\\
The total molecular wavefunction can be expressed in the so-called Born-Huang representation~\cite{bornhuang},
\begin{equation}
\Psi(\bs r, \bs R, t) = \sum_l^\infty \chi_{BO}^{(l)}(\bs R, t) \Phi^{(l)}_{\bs R}(\bs r) \, ,
\label{bh}
\end{equation}
where $\Phi^{(l)}_{\bs R}(\bs r)$ are solutions of the time-independent electronic Schr\"{o}dinger equation with corresponding electronic energy $\epsilon_{BO}^{(l)}(\bs R)$,
\begin{equation}
\hat{H}_{BO}(\bs r, \bs R) \Phi^{(l)}_{\bs R}(\bs r) = \epsilon_{BO}^{(l)}(\bs R) \Phi^{(l)}_{\bs R}(\bs r)\, .
\end{equation}
 $\chi_{BO}^{(l)}(\bs R, t)$ represents a BO nuclear contribution in electronic state $(l)$.

\begin{figure}
\center
\includegraphics[width=1.0\textwidth]{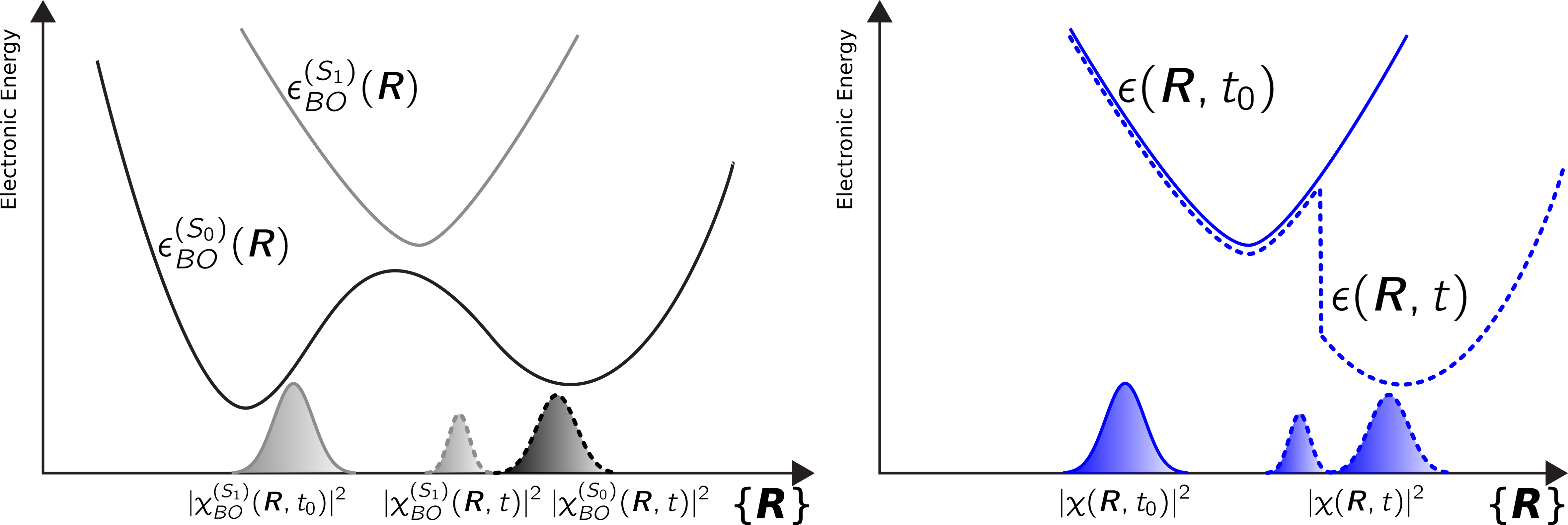}
\caption{Schematic comparison between the Born-Huang (left panel) and Exact Factorization (right panel) representation of a nonadiabatic event. In both cases, a system initiated a time t$_0$ on $\epsilon_{BO}^{(S_1)}$ (continuous line) is evolved until a given time $t$, after the nonadiabatic event (dashed lines). In the Born-Huang representation, the initial nuclear wavefunction in $S_1$ (in gray) will be split at time $t$ into a contribution in $S_0$ (in black) and a smaller one remaining in $S_1$ (in gray). In the Exact Factorization picture, only a single time-dependent nuclear wavefunction is present. The potential energy surface on which this wavefunction evolves will vary in time. At time $t_0$ (continuous blue curve), the TDPES shape resembles $\epsilon_{BO}^{(S_1)}$, while once the nonadiabatic region is passed, the TDPES exhibits a step connecting two pieces of BO surfaces (dashed blue curve, down-shifted for clarity).}
\label{fig0}
\end{figure}

The Born-Huang representation is a common starting point for a plethora of techniques aiming at solving the coupled electron-nuclear dynamics for molecular systems (for reviews, see Ref~\cite{marxbook,barbatti10a,curchod2013trajectory,malhado2014non}). By using solutions of the time-independent electronic Schr\"{o}dinger equation to express the total molecular wavefunction, the Born-Huang representation leads to the concept of nuclear wavefunctions evolving on time-independent potential energy surfaces (left panel of Fig.~\ref{fig0}).
\\
In the following, we will preserve the time-dependence in both the electronic and the nuclear contributions to the full molecular wavefunction, leading to a representation where a \textit{single} product can be used. The \textit{exact factorization} (EF) of the molecular wavefunction is defined by
\begin{equation}
\Psi(\bs r, \bs R, t) = \chi(\bs R, t) \Phi_{\bs R}(\bs r, t)\, ,
\label{ef}
\end{equation}
with the partial normalization condition $\int d\bs r |\Phi_{\bs R}(\bs r, t)|^2=1$, $\forall \bs R,t$.
Note the central difference between Eqs.~\eqref{ef} and~\eqref{bh}: the summation over the electronic states is replaced by a time-dependence in the electronic wavefunction.
\\
The partial normalization condition guarantees that the squared modulus of the nuclear wavefunction is the exact nuclear density, computed from the molecular wavefunction. The relation
\begin{align}\label{eqn: nuclear density}
\left|\chi(\bs R,t)\right|^2 = \int d\bs r\left|\Psi(\bs r,\bs R,t)\right|^2= \sum_l^\infty \left|\chi_{BO}^{(l)}(\bs R, t) \right|^2,
\end{align}
between the (exact) nuclear wavefunction and the BO wavefunctions, then follows. Furthermore, representing the electronic wavefunction $\Phi_{\bs R}(\bs r, t)$ in terms of BO electronic states, together with Eq.~\eqref{eqn: nuclear density}, yields
\begin{align}\label{eqn: product vs sum}
\Psi(\bs r, \bs R,t) = \chi(\bs R, t) \Phi_{\bs R}(\bs r, t) = e^{\frac{i}{\hbar}S(\bs R,t)}\sqrt{\sum_l^\infty \left|\chi_{BO}^{(l)}(\bs R, t) \right|^2} \left(\sum_l^\infty C_{l}(\bs R, t) \Phi^{(l)}_{\bs R}(\bs r)\right).
\end{align}
This expression clearly shows that a single-product form can be used for the molecular wavefunction and is not in disagreement with the Born-Huang representation of Eq.~\eqref{bh}. In Eq.~\eqref{eqn: product vs sum} we have used the symbol $C_{l}(\bs R, t)$ for the expansion coefficients of the electronic wavefunction and $S(\bs R,t)$ is a nuclear phase factor that will be discussed below.

The equations of motion for the electronic -- $\Phi_{\bs R}(\bs r, t)$ --  and the nuclear -- $\chi(\bs R, t)$ -- contribution to the total molecular wavefunction read
\begin{equation}
\left(\hat{H}_{BO}(\bs r, \bs R) + \hat{U}_{en}^{coup}[\Phi_{\bs R},\chi]  -\epsilon(\bs R,t)\right) \Phi_{\bs R}(\bs r, t) = i\hbar \partial_t  \Phi_{\bs R}(\bs r, t) 
\label{eqel}
\end{equation}
\begin{equation}
\left(\sum_{\nu=1}^{N_n} \frac{\left[ -i\hbar \nabla_\nu + \bs A_\nu(\bs R,t)\right]^2}{2M_\nu} + \epsilon(\bs R,t) \right) \chi(\bs R, t)= i\hbar \partial_t \chi(\bs R, t) \, .
\label{eqnuc}
\end{equation}
Detailed derivation and discussions on these equations were presented in the literature~\cite{Gross_PRL2010,Gross_JCP2012,e16010062}, and the interested reader is particularly referred to the complete analysis proposed in Ref.~\cite{Gross_JCP2015}. $\hat{U}_{en}^{coup}[\Phi_{\bs R},\chi]$ corresponds to an electron-nuclear coupling operator and is defined by
\begin{equation}
\hat{U}_{en}^{coup}[\Phi_{\bs R},\chi]=\sum_{\nu=1}^{N_n} \frac{1}{M_\nu} \left[ \frac{\left[ -i\hbar \nabla_\nu - \bs A_\nu(\bs R,t)\right]^2}{2} 
+\left(\frac{-i\hbar \nabla_\nu \chi(\bs R,t)}{\chi(\bs R,t)} + \bs A_\nu(\bs R,t) \right)\left( -i\hbar \nabla_\nu - \bs A_\nu(\bs R,t) \right)\right]\, ,
\end{equation}
with $\bs A_{\nu}(\bs R,t)=\langle \Phi_{\bs R}(t) | -i\hbar\nabla_{\nu} \Phi_{\bs R}(t) \rangle_{\bs r}$ being a time-dependent vector potential.
\\
More important in the context of this work, the equation of motion for the nuclear wavefunction -- Eq.~\eqref{eqnuc} -- contains a time-dependent scalar potential termed ``time-dependent potential energy surface'' (TDPES), symbolized by $\epsilon(\bs R,t)$ and defined as
\begin{align}
\epsilon(\bs R,t) &=\epsilon_{GI,1}(\bs R,t)+\epsilon_{GI,2}(\bs R,t)+\epsilon_{GD}(\bs R,t) \notag \\
&= \langle \Phi_{\bs R}(t) | \hat{H}_{BO} + \hat{U}_{en}^{coup} -i\hbar\partial_t | \Phi_{\bs R}(t) \rangle_{\bs r}\, .
\label{eqtdpes}
\end{align}
The subscripts ``GI'' and ``GD'' stand for the \textit{gauge-independent} and \textit{gauge-dependent} contributions to the total TDPES. The exact factorization in Eq.~\eqref{ef} is indeed unique up to a gauge transformation~\cite{Gross_JCP2015}. In the following sections, the time-dependent vector potential $\bs A_{\nu}(\bs R,t)$ will be set to zero, fixing the gauge freedom and transferring all the electronic backreaction to the TDPES~\cite{Gross_JCP2015}. The nuclear wavefunction can be expressed in a polar form $\chi(\bs R,t)=|\chi(\bs R,t)|e^{iS(\bs R,t)/\hbar}$ and, for 1D problem, imposing to the phase $S(R,t)$ the condition
\begin{equation}
S(R,t)=\int^R dR'\frac{\Im \left[\langle \Psi(t) | \partial_{R'} \Psi(t) \rangle_{r}\right]}{|\chi(R',t)|^2}\label{fixing the gauge}
\end{equation}
leads to $A(R,t)=0$~\cite{Gross_JCP2015,Min_PRL2014,Requist_PRA2015}.\\
When a nuclear wavefunction evolves solely in a given electronic state, the corresponding TDPES resembles its adiabatic BO surface (continuous curve, right panel of Fig.~\ref{fig0}). During a passage through a nonadiabatic region, the TDPES follows diabatic curves, connecting smoothly the two involved adiabatic BO states. After the nuclear wavefunction splitting and far from the crossing region, the TDPES will display a step that connects two piecewise adiabatic surfaces (dashed curve, right panel of Fig.~\ref{fig0}). If this overall dynamical behavior of the TDPES applies to cases where BO nuclear wavefunctions interfere in the nonadiabatic regions is the central theme of the present work.

\subsection{Nonadiabatic quantum interferences}\label{interference}
\begin{figure}
\center
\includegraphics[width=1.0\textwidth]{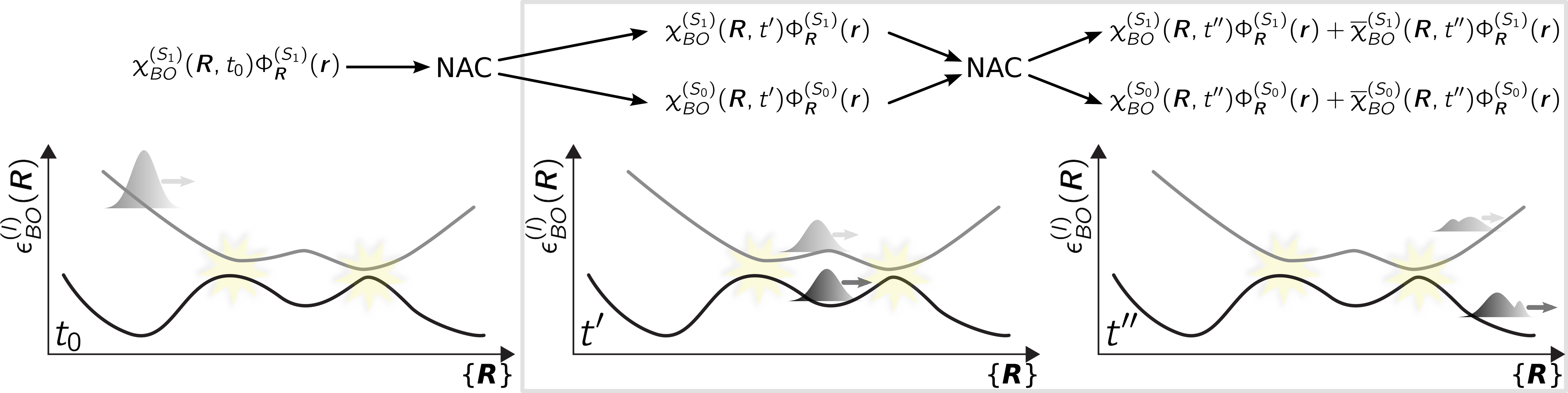}
\caption{Appearance of interference effects in nonadiabatic dynamics. Time $t_0$: initial condition for the nonadiabatic dynamics; time $t'$: after the first nonadiabatic crossing (NAC); time $t''$: after the second NAC. The events of interest for the present work are framed by the gray rectangle.}
\label{fig0bis}
\end{figure}

In order to define quantum interferences in the context of nonadiabatic dynamics, let us consider the following situation (schematically depicted in Fig.~\ref{fig0bis}).

The nuclear wavefunction is initiated in $S_1$, as shown in Fig.~\ref{fig0bis} at time $t_0$, and undergoes a first nonadiabatic event leading to its branching into two contributions, one that keeps travelling on the $S_1$ surface, the other being produced on $S_0$ (Fig.~\ref{fig0bis} at time $t'$). The molecular wavefunction, initially the product $\Psi(\bs r,\bs R,t_0)=\chi_{BO}^{(S_1)}(\bs R,t_0)\Phi_{\bs R}^{(S_1)}(\bs r)$, becomes a superposition of the contributions in each electronic state $\Psi(\bs r,\bs R,t')=\chi_{BO}^{(S_0)}(\bs R,t')\Phi_{\bs R}^{(S_0)}(\bs r)+\chi_{BO}^{(S_1)}(\bs R,t')\Phi_{\bs R}^{(S_1)}(\bs r)$. The two BO nuclear wavefunctions evolve on the corresponding adiabatic surfaces, feeling -- in the most general case -- different forces. While the two wavefunctions might separate in configuration space and never meet again, there exists an interesting case where they both reach a (common) second nonadiabatic region at a later time. At this point, and without loss of generality, the full molecular wavefunction can be expressed as
\begin{align}\label{eqn: Psi at time t''}
\Psi(\bs r,\bs R,t'')=\left(\chi_{BO}^{(S_0)}(\bs R,t'')+\overline\chi_{BO}^{(S_0)}(\bs R,t'')\right)\Phi_{\bs R}^{(S_0)}(\bs r)+\left(\chi_{BO}^{(S_1)}(\bs R,t'')+\overline\chi_{BO}^{(S_1)}(\bs R,t'')\right)\Phi_{\bs R}^{(S_1)}(\bs r),
\end{align}
as shown in Fig.~\ref{fig0bis} at time $t''$. Similarly to time $t'$, $\Psi(\bs r,\bs R,t'')$ is a superposition of the two BO contributions from $S_0$ and $S_1$. However, the notation indicates here that $\chi_{BO}^{(S_0)}(\bs R,t'')$ and $\overline\chi_{BO}^{(S_1)}(\bs R,t'')$ have been generated by $\chi_{BO}^{(S_0)}(\bs R,t')$  during the second nonadiabatic event, and the same idea applies to $\chi_{BO}^{(S_1)}(\bs R,t'')$ and $\overline\chi_{BO}^{(S_0)}(\bs R,t'')$.

The different histories of the nuclear wavefunction dynamics can result in specific recombination patterns. \textsl{Quantum interferences} will be observed whenever a non-zero overlap exists in the second nonadiabatic region between the BO nuclear wavefunction contributions. In the expression of the total nuclear density,  
\begin{align}
\left|\chi(\bs R,t'')\right|^2=\left|\chi_{BO}^{(S_0)}(\bs R,t'')\right|^2 + \left|\overline\chi_{BO}^{(S_0)}(\bs R,t'')\right|^2 +
\left|\chi_{BO}^{(S_1)}(\bs R,t'')\right|^2 + \left|\overline\chi_{BO}^{(S_1)}(\bs R,t'')\right|^2\label{eqn: nuclear density with interference}\\
+2\Re\left[\left({\chi_{BO}^{(S_0)}}(\bs R,t'')\right)^*\overline\chi_{BO}^{(S_0)}(\bs R,t'')+\left({\chi_{BO}^{(S_1)}}(\bs R,t'')\right)^*\overline\chi_{BO}^{(S_1)}(\bs R,t'')\right],\nonumber 
\end{align}
interferences can be identified in the last term on the right-hand-side. They are interferences produced by BO wavefunctions corresponding to the same electronic state, but ``generated'' by wavefunctions with different histories, e.g., $\chi_{BO}^{(S_0)}(\bs R,t'')$ is generated by $\chi_{BO}^{(S_0)}(\bs R,t')$ while $\overline\chi_{BO}^{(S_0)}(\bs R,t'')$ is generated by $\chi_{BO}^{(S_1)}(\bs R,t')$.
\\
Quantum interferences are affected by various factors, e.g., initial conditions or different histories of the BO wavefunctions before recrossing. Hence, their effect on the final electronic populations resulting from nonadiabatic transitions can be strong~\cite{granucci1995coherent,romstad1997nonadiabatic,donoso2000simulation}, and this effect is often termed ``St\"{u}ckelberg oscillations''~\cite{tully90,shevchenko2010landau}. 

Independent crossings will instead occur if the BO nuclear wavefunctions fully separate in configuration space before the second nonadiabatic region. In this case the interference terms in Eq.~\eqref{eqn: nuclear density with interference} are zero and the nuclear density is simply the sum of independent contributions.

The analysis we propose in the following sections aims at identifying signatures of quantum interferences in the EF framework. In this context, the concept of multiple BO nuclear wavefunctions, $\chi_{BO}^{(S_0)}$ and $\chi_{BO}^{(S_1)}$, is replaced by a single nuclear wavefunction, $\chi$, but, as shown in Eq.~\eqref{eqn: nuclear density}, both formulations lead to the same nuclear density. Therefore, rather than rationalizing the appearance of quantum interferences in terms of BO wavefunctions (second line on the right-hand-side of Eq.~\eqref{eqn: nuclear density with interference}), we analyze the properties of the TDPES, the very quantity responsible for the nuclear wavefunction dynamics.

\section{Model system}
\label{model}

\subsection{Proton-coupled electron transfer model}
The nonadiabatic quantum dynamics investigated in the present work is based on the 1D Shin-Metiu model~\cite{shin1995nonadiabatic} {\color{black}for nonadiabatic electron transfer. The system consists of three ions and a single electron, as depicted in Fig.~\ref{fig: metiu model}.
\begin{figure}[h!]
 \centering
 \includegraphics*[width=.4\textwidth]{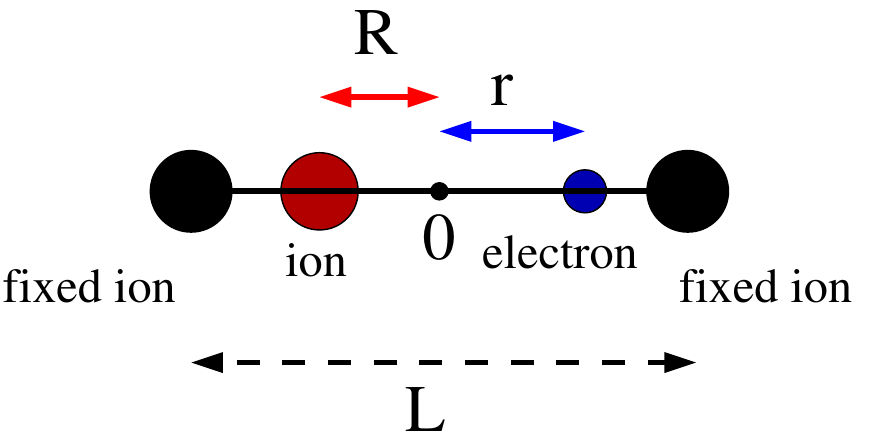}
 \caption{Schematic representation of the model system described by the Hamiltonian in Eq.~\eqref{eqn: metiu-hamiltonian}. $R$ and $r$ indicate the coordinates of the moving ion and electron, respectively, in one dimension. $L$ is the distance between the fixed ions.}
 \label{fig: metiu model}
\end{figure}
Two ions are fixed at a distance of $L=19.05$~a.u., the third ion and the electron are free to move in one dimension along the line joining the two fixed ions. The Hamiltonian of this system reads
\begin{align}
 \hat{H}(r,R)= -\frac{1}{2}\frac{\partial^2}{\partial r^2}-\frac{1}{2M}\frac{\partial^2}{\partial R^2}  +
 \frac{1}{\left|\frac{L}{2}-R\right|}+\frac{1}{\left|\frac{L}{2} + R\right|}
 -\frac{\mathrm{erf}\left(\frac{\left|R-r\right|}{R_f}\right)}{\left|R - r\right|}
 -\frac{\mathrm{erf}\left(\frac{\left|r-\frac{L}{2}\right|}{R_r}\right)}{\left|r-\frac{L}{2}\right|}
 -\frac{\mathrm{erf}\left(\frac{\left|r+\frac{L}{2}\right|}{R_l}\right)}{\left|r+\frac{L}{2}\right|}.\label{eqn: metiu-hamiltonian}
\end{align}
Here, the symbols $\bs r$ and $\bs R$ are replaced by $r$ and $R$, the coordinates of the electron and the movable ion measured from the center of the two fixed ions. As it can be seen from Eq.~\eqref{eqn: metiu-hamiltonian}, the movable ion interacts with the fixed ions via a bare Coulomb potential, whereas the electron interacts with all ions via a soft-Coulomb potential. The ionic mass is chosen as $M=1836$, the proton mass, whereas the other parameters are tuned in order to make the system essentially a two-electronic-state model.} We used the following parameters for the model: $R_r=4.0$ a.u., $R_f=4.5$ a.u., and $R_l=2.9$ a.u.. This parameter set allows to focus on the first two Born-Oppenheimer potential energy curves ($\epsilon_{BO}^{(S_0)}(R)$ and $\epsilon_{BO}^{(S_1)}(R)$ in the inset of Fig.~\ref{fig1}), producing a two-state model system. Interaction between the two electronic states leads to a strong region of nonadiabaticity in the region around $R=-2.0$ a.u.. Both $\epsilon_{BO}^{(S_1)}(R)$ and $\epsilon_{BO}^{(S_0)}(R)$ exhibit a similar shape before the nonadiabatic region, with a slightly smaller slope for the latter in the direct vicinity of the coupling. On the other hand, the shape of the two PESs strongly differ towards higher $R$ values ($< 4.0$ a.u.). 

\subsection{Initial conditions and computational details}
\label{icandcd}
We are interested in studying the relation between interferences and the shape of the TDPES in nondiabatic events. As discussed in Sec.~\ref{interference}, a way to carefully control the interferences in the nonadiabatic region is to initiate the quantum dynamics with a molecular wavefunction being described by a superposition of two slightly different wavefunctions (to reconstruct the situation at time $t'$ of Fig.~\ref{fig0bis}). Similar initial conditions were used in the past to study wavepacket interferometry~\cite{granucci1995coherent,romstad1997nonadiabatic,donoso2000simulation}. This initial molecular wavefunction mimics the important situation observed in multiple nonadiabatic crossings, i.e., when two nuclear wavepackets formed after a first nonadiabatic event meet at a later time in a coupling region. As the two contributions undergo different dynamics before reaching the coupling region, their recombination can lead to quantum interferences.  

The initial wavefunction in our simulations is defined as:
\begin{equation}
\Psi(r, R, t_0) = \chi_{BO}^{(S_0)}(R, t_0) \Phi^{(S_0)}_{ R}( r)  + \chi_{BO}^{(S_1)}(R, t_0) \Phi^{(S_1)}_{R}(r), 
\label{iniwf}
\end{equation}
where $\chi_{BO}^{(S_0)}(R, t_0) = N\exp\left(-\frac{(R-R_0)^2}{2\sigma^2} + \frac{i}{\hbar} P_{in,S_0}(R-R_0) \right)$ and $\chi_{BO}^{(S_1)}(R, t_0) = N\exp\left(-\frac{(R-R'_0)^2}{2\sigma^2}\right)$. We used for all simulations $\sigma=0.5$~a.u., $R_0=-6.0$~a.u., and $R'_0=-4.0$~a.u.. The normalization is set such that $\int dR |\chi_{BO}^{(S_0)}(R, t_0)|^2= \int dR |\chi_{BO}^{(S_1)}(R, t_0)|^2=0.5$.  Such initial condition therefore describes a superposition between two nuclear wavefunction contributions (one in each electronic state considered). The initial wavefunction in $S_1$ has no initial momentum. The other contribution, in $S_0$, has an initial momentum $P_{in,S_0}$ that will be tuned in the following numerical simulations. Altering the dynamics of the initial $S_0$ contribution will indeed result in different quantum interference effects in the coupling region, and potentially to a different final population of the electronic states.
\\
In Sec.~\ref{interference} we have briefly discussed also the case of independent crossings, leading to the absence of interferences in the coupling region. In order to reproduce this situation in the studied model and, thus, to clarify the role of interference effects in the nonadiabatic region, we have performed additional quantum dynamics simulations with the following initial conditions:
\begin{equation}
\tilde{\Psi}_1(r, R, t_0) = \tilde{\chi}_{BO}^{(S_0)}(R, t_0) \Phi^{(S_0)}_{ R}( r) + 0 \cdot \Phi^{(S_1)}_{ R}( r)
\label{iniwf1}
\end{equation}
and
\begin{equation}
\tilde{\Psi}_2(r, R, t_0) = 0 \cdot \Phi^{(S_0)}_{ R}( r) +  \tilde{\chi}_{BO}^{(S_1)}(R, t_0) \Phi^{(S_1)}_{ R}( r) \, ,
\label{iniwf2}
\end{equation}
where $\tilde{\chi}_{BO}^{(S_0)}(R, t_0)\propto\chi_{BO}^{(S_0)}(R, t_0)$ and $\tilde{\chi}_{BO}^{(S_1)}(R, t_0) \propto\chi_{BO}^{(S_1)}(R, t_0)$. Here, only the normalization constant is modified with respect to Eq.~\eqref{iniwf}, as the BO nuclear probability densities integrate to one. These simulations will be termed ``IWA'' (independent wavefunction approximation) in the following, as they represent the ideal case of independent, i.e., non-interfering, dynamics for each initial contribution given in Eq.~\eqref{iniwf}. 

In addition, we performed standard TSH calculations as described in Ref.~\cite{tully90}, using a total of 2000 classical trajectories. The initial conditions were sampled from the Wigner distribution of $\chi_{BO}^{(S_0)}(R, t_0)$ for 1000 trajectories, and of $\chi_{BO}^{(S_1)}(R, t_0)$ for the other 1000 classical trajectories. All amplitudes for the TSH trajectories were initiated in the same way: $C_0^{TSH}(t_0)=C_1^{TSH}(t_0)=\sqrt{0.5}+0.0i$. 

The TDPES is obtained for each simulation by {\color{black}first} computing the full wavefunction $\Psi(r,R,t)$ for the system, {\color{black}then using the definition given in Eq.~\eqref{eqtdpes}. To this end, we determine the electronic wavefunction as $\Phi_R(r,t)=\Psi(r,R,t)/(e^{(i/\hbar)S(R,t)}|\chi(R,t)|)$. Here, the phase $S(R,t)$ of the nuclear wavefunction is given by the gauge condition of Eq.~\eqref{fixing the gauge}, whereas the modulus $|\chi(R,t)|$ is given by the exact nuclear density, which is available via integration over the electronic coordinate of $|\Psi(r,R,t)|^2$. We refer to the TDPES thus computed as ``exact potential'', since it is determined from the (numerically) exact solution of the full time-dependent Schr\"odinger equation, and no information about the adiabatic electronic states is used to determine its form.}

All quantum dynamics calculations are performed using the split-operator technique~\cite{spo}, {\color{black} to solve numerically the full time-dependent Schr\"odinger equation for $\Psi(r,R,t)$,} using an integration time step of $dt=0.1$~a.u.; classical trajectories evolved on the exact TDPES are integrated with the velocity-Verlet algorithm with $dt=0.1/0.5$~a.u. and the initial conditions are sampled as described above for the TSH calculations; in TSH dynamics $dt=0.01$~a.u., classical trajectories are integrated with the velocity-Verlet algorithm and the fourth-order Runge-Kutta algorithm is used to solve the electronic equation. {\color{black}It is worth noting that since in the quantum dynamics simulations we propagate numerically the full electron-nuclear wavefunction $\Psi(r,R,t)$ according to the Hamiltonian of Eq.~\eqref{eqn: metiu-hamiltonian}, with the initial conditions discussed above, the whole manifold of electronic adiabatic states is automatically included. We will use two adiabatic electronic states only for the analysis performed in the following sections, as the populations of electronic states that are higher in energy are negligible (three or more orders of magnitude smaller than the populations of states $S_0$ and $S_1$). Consequently, TSH calculations are performed in the restricted space of two electronic states. We have checked that the results presented below are not affected if the state $S_2$ is included in the calculations.}

\section{Results and discussion}
\label{results}

\subsection{Shape of the TDPES in the event of nonadiabatic quantum interferences}
\label{tdpesshape}
\begin{figure}
\center
\includegraphics[width=0.9\textwidth]{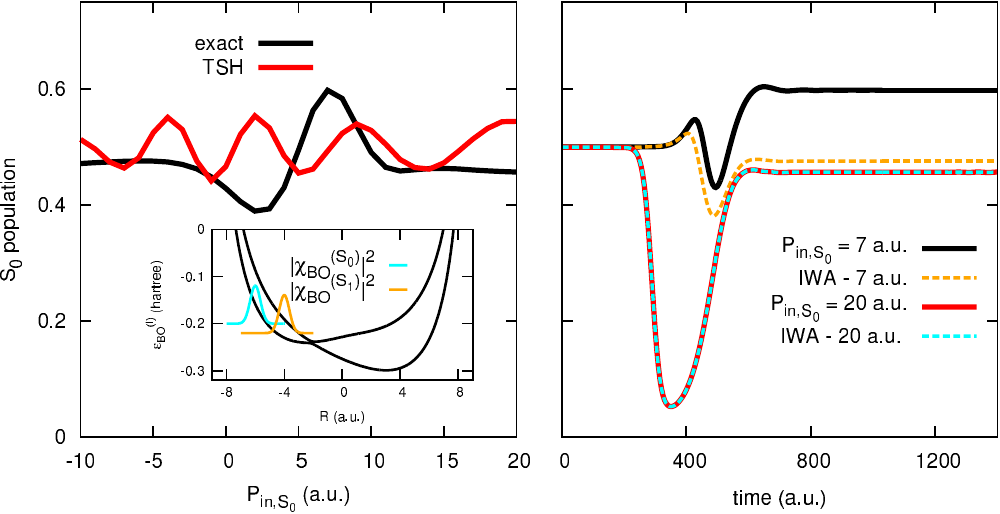}
\caption{Left panel: Population in $S_0$ for different initial momenta of the $S_0$ wavefunction, $P_{in,S_0}$. The result obtained from an exact nuclear dynamics is compared with that of TSH. The overall probability density for the initial wavefunction $|\chi(R,t_0)|^2$ is composed of a $S_0$ ($|\chi_{BO}^{(S_0)}(R,t_0)|^2$) and a $S_1$ contribution ($|\chi_{BO}^{(S_1)}(R,t_0)|^2$), as depicted in the inset and superimposed on the corresponding BO potential energy curves ($\epsilon_{BO}^{(S_0)}$ and $\epsilon_{BO}^{(S_1)}$).
Right panel: $S_0$ population trace for the $P_{in,S_0}= 7$~a.u. and 20~a.u. runs. The exact nuclear dynamics (continuous lines) is compared with the IWA (dashed lines).}
\label{fig1}
\end{figure}

As mentioned in Sec.~\ref{model}, the system under consideration in this work is designed to study the EF description of interferences due to interactions between two nuclear wavefunctions in a nonadiabatic region (each nuclear wavefunction initially evolves in a specific BO state). We monitored the final $S_0$ population with respect to the initial momentum of the $S_0$ contribution (Fig.~\ref{fig1}, left panel) -- the $S_1$ contribution being always the same, as described in Eq.~\eqref{iniwf}. The overall $S_0$ population with respect to $P_{in,S_0}$ does not show particular features, except in the region where $P_{in,S_0}$ is comprised between -2 and 12 a.u.. In this particular range of initial momenta, oscillations in the final population are observed and can be attributed to quantum interferences between wavefunction components in the coupling region altering the final populations~\cite{donoso2000simulation}, as detailed below. 
TSH is in qualitatively good agreement with the average final population of the exact result, even though it misses the aforementioned oscillation between -2 and 12 a.u.. The oscillatory behavior of the TSH result can be related to the overcoherent superposition of its complex amplitudes, as described in details in different recent works~\cite{subotnik2011decoherence}, while the interference feature observed in the exact calculation is not clearly reproduced due to the independent classical trajectory approximation. 
\\
What happens if each initial component of the total wavefunction is run independently? Such simulations -- the IWA mentioned in Sec.~\ref{icandcd} -- would surely highlight the role of mutual interferences during the nonadiabatic process, since in this case the last term in Eq.~\eqref{eqn: nuclear density with interference} is not present. When applied to an initial $S_0$ momentum leading to weak interferences in the coupling region (for example $P_{in,S_0}=20$~a.u.), the summed $S_0$ population during the IWA dynamics leads to an excellent agreement with the $S_0$ population time trace for the exact dynamics (as depicted in the left panel of Fig.~\ref{fig1}). In contrast, the IWA fails to reproduce the exact $S_0$ population when interference effects are important ($P_{in,S_0}=7$~a.u.), and in this sense mirrors the behavior of TSH. The IWA will therefore be used as a tool in the following to shed light on the relation between interferences and the dynamical shape of the TDPES.

\begin{figure}
\center
\includegraphics[width=1.0\textwidth]{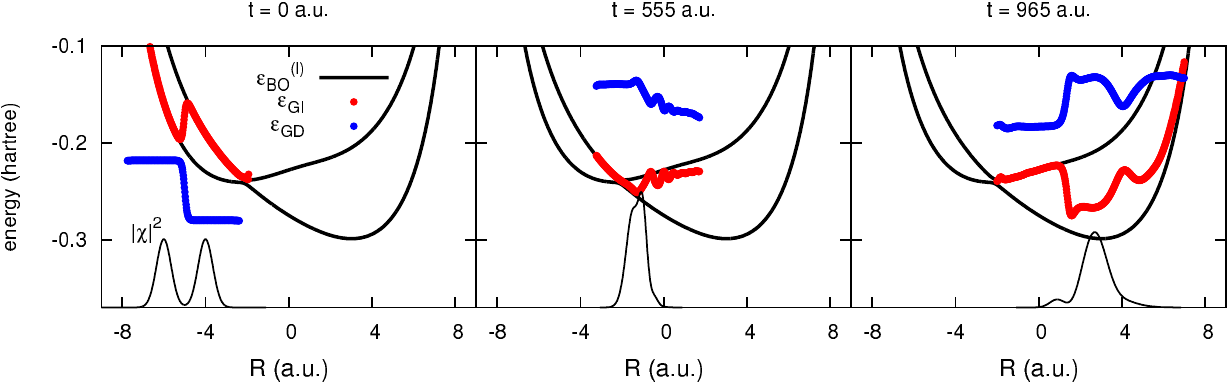}
\caption{Snapshots at $t=$0, 555, and 965~a.u. of the GI (red) and GD (blue) contributions to the TDPES, during the dynamics with $P_{in,S_0}= 7$~a.u.. BO potential energy surfaces ($\epsilon_{BO}^{(l)}$) are indicated with thick black curves, and the nuclear probability density distribution $|\chi(R,t)|^2$ is depicted with a thin black line.} 
\label{fig2}
\end{figure}

We first start our analysis of the TDPES by observing its overall behavior during a nonadabatic dynamics containing quantum interferences effects. Snapshots of the TDPES at three different times for the run with $P_{in,S_0}=7$~a.u. are depicted in Fig.~\ref{fig2}. At $t=0$~a.u., the initial condition leads to a step in the GI part of the TDPES ($\epsilon_{GI,1}(R, t)+\epsilon_{GI,2}(R, t)$ in Eq.~\eqref{eqtdpes}) that bridges the $S_0$ and $S_1$ adiabatic surfaces. At the initial time, the GD part of the TDPES also presents a step between two constant values, mirroring the step in the GI part. Soon after the passage of the system through the nonadiabatic region ($t=555$~a.u., middle panel of Fig.~\ref{fig2}), the TDPES shape shows oscillations, and its description in terms of adiabatic surfaces connected by simple steps is no more possible. It is important to note, however, that the GD contribution to the TDPES also in this case mirrors the GI one, as observed for simple nonadiabatic events~\cite{Gross_MP2013}. At a later time, $\epsilon_{GI}(R, t)$ reflects the complex composition of the total wavefunction, and its shape matches in some regions of the configuration space the shape of a given BO surface. For example, the left shoulder ($0.0<R<1.0$~a.u.) of the overall nuclear probability density distribution belongs, in a BO picture, to the $S_1$ state. Hence, it is not surprising that the TDPES in this particular region resembles $\epsilon_{BO}^{(S_1)}$. In other regions, the TDPES seems to acquire a mean-field character, mixing variable contributions of $\epsilon_{BO}^{(S_0)}$ and $\epsilon_{BO}^{(S_1)}$.

\subsection{Analysis of the TDPES for weak and strong nonadiabatic quantum interference effects}
\label{tdpesanalysis}
The first observations presented in Sec.~\ref{tdpesshape} suggest that the TDPES develops a rather complex structure whenever nonadiabatic quantum interferences take place in the dynamics. To shed more light on this intricate structure, the full TDPES should be monitored for cases with either strong ($P_{in,S_0}= 7$~a.u.) or weak ($P_{in,S_0}= 20$~a.u.) interference effects. Yet we need first to determine a critical time in the dynamics to perform such analysis. As discussed in Sec.~\ref{interference}, quantum interferences are observed when the BO nuclear wavefunctions arrive concurrently in the nonadiabatic region, leading to non-zero overlaps between transmitted and transferred BO wavefunctions. It seems therefore natural to introduce the following indicator: 
\begin{equation}
\eta(t)= \left|\int_\Delta dR \left(\chi_{BO}^{(S_0)}(R,t)\right)^* \chi_{BO}^{(S_1)}(R,t) \right|=\left|\int_\Delta dR |\chi(R,t)|^2 \left(C_{S_0}(R,t)\right)^*C_{S_1}(R,t) \right|  \, ,
\label{eqindic}
\end{equation}
which monitors the overlap of the BO wavefunctions in the region $\Delta$ as a function of time (first term on the right-hand-side). Here, $\Delta=\lbrace R: -3~\text{a.u.}\leq R\leq -1~\text{a.u.}\rbrace$. In the EF framework, however, we do not have direct access to $\chi_{BO}^{(l)}(R,t)$, rather to $\chi(R,t)$ and to the projections of the electronic wavefunction on the adiabatic states, i.e., the coefficients $C_{l}(R,t)=\chi_{BO}^{(l)}(R,t)/\chi(R,t)$ introduced in Eq.~\eqref{eqn: product vs sum}. We have therefore expressed the overlap in terms of quantities defined in the context of the EF (second term on the right-hand-side).

The behavior of $\eta(t)$ as a function of time allows us to identify a time interval for the crossing of the nonadiabatic region where quantum interferences potentially appear, since the BO wavefunctions have a non-zero overlap. In particular, we expect that when independent crossings take place, $\eta(t)$ will have a well-defined double-peak shape: one peak will correspond to the first BO wavefunction entering the coupling region, transferring amplitude on the other state and then leaving the coupling region; the second peak will instead correspond to the second BO wavefunction following an analogous process. We also expect that no (or small) difference will be observed between the exact dynamics and the IWA during independent crossings.

\begin{figure}
\center
\includegraphics[width=0.8\textwidth]{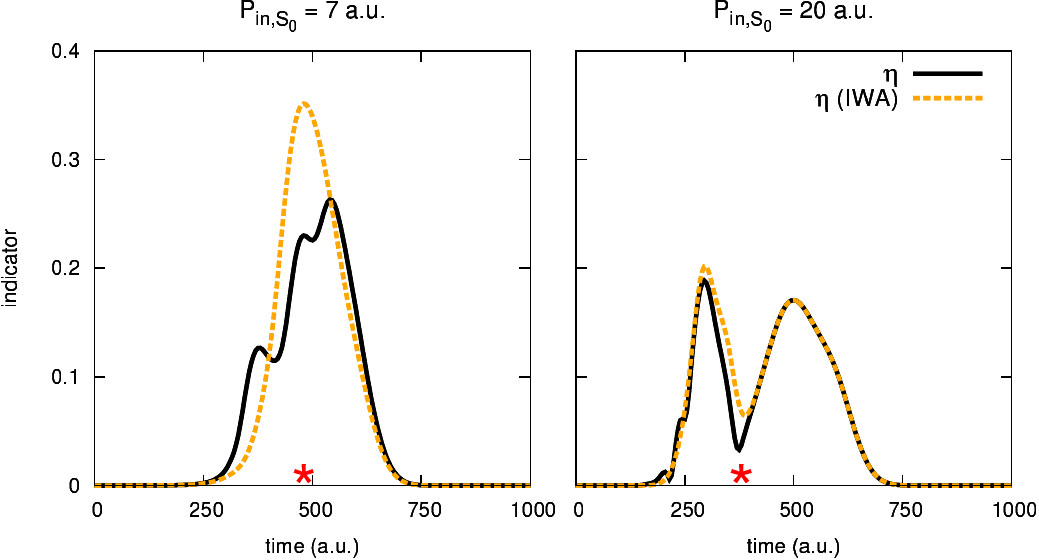}
\caption{Indicator $\eta$ during the quantum dynamics initiated with $P_{in,S_0}= 7$~a.u. (left) and $P_{in,S_0}= 20$~a.u. (right). The indicator computed for the full dynamics is given in black, while the indicator for the quantum dynamics within the IWA is depicted with a dashed line. Asterisks indicate a specific time used for further analysis.} 
\label{fig3}
\end{figure}

\begin{figure}
\center
\includegraphics[width=0.8\textwidth]{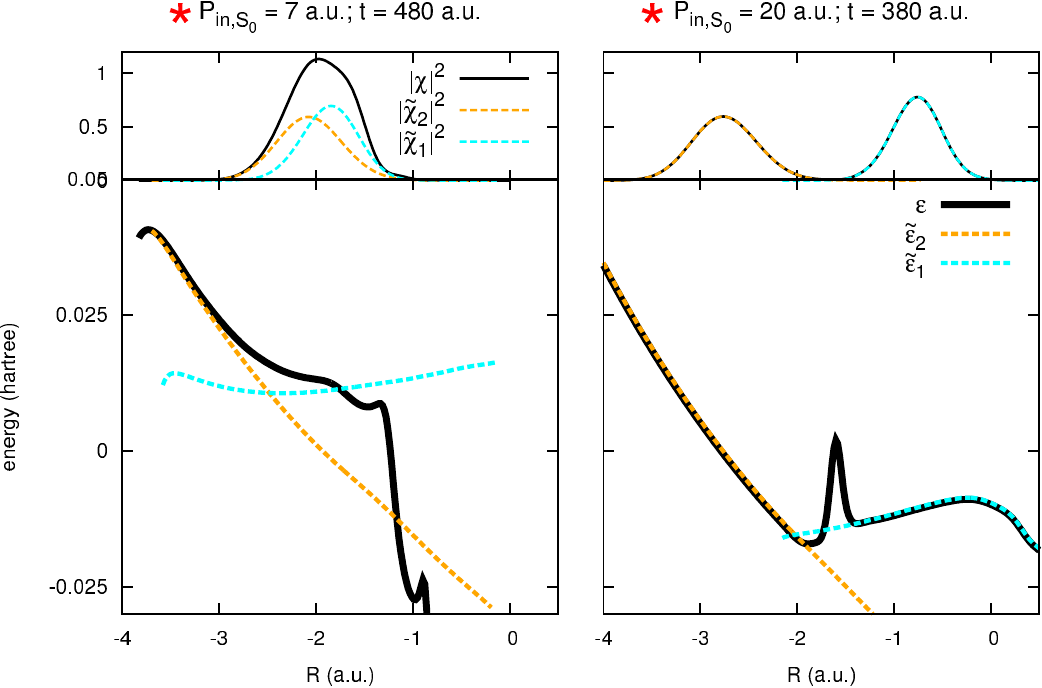}
\caption{Analysis of the full TDPES for selected times in the $P_{in,S_0}= 7$~a.u. (left) and $P_{in,S_0}= 20$~a.u. (right) runs (see red asterisks in Fig.~\ref{fig3}). The TDPES for the full ($\epsilon$, black curve) and the IWA dynamics ($\tilde{\epsilon}_1$ and $\tilde{\epsilon}_2$, cyan and orange dashed curve, respectively) are compared. Dashed curves were rigidly $y$-shifted for clarity. We note that $\tilde{\chi}_1(R,t)$ and $\tilde{\chi}_2(R,t)$ do \textit{not} represent the exact $S_0$ and $S_1$ contributions to the full nuclear wavefunction, but are instead defined as $|\tilde{\chi}_1(R,t)|^2= \int dr |\tilde{\Psi}_1(r, R, t)|^2$ and $|\tilde{\chi}_2(R,t)|^2= \int dr |\tilde{\Psi}_2(r, R, t)|^2$. Upper panels give a snapshot of the exact ($|\chi(R,t)|^2$) and IWA ($|\tilde{\chi}_1(R,t)|^2$ and $|\tilde{\chi}_2(R,t)|^2$) probability densities.} 
\label{fig4}
\end{figure}

We have monitored $\eta(t)$ for both the exact dynamics and the IWA to highlight specific times where interferences play a role in the $P_{in,S_0}= 7$~a.u. dynamics (left panel of Fig.~\ref{fig3}). While $\eta(\text{IWA})$ only peaks in the coupling region with no additional structure, $\eta$ for the full dynamics displays oscillations. We investigated the shape of the TDPES for a time in this particular dynamics where $\eta(\text{IWA})$ shows a maximum, while $\eta$ has a marked structure (red asterisk at $t=480$~a.u. in left panel of Fig.~\ref{fig3}). As anticipated above, the dynamics characterized by $P_{in,S_0}= 20$~a.u. does not lead to important deviations between the different initial conditions, $\eta$ being close to $\eta(\text{IWA})$ except for a small period of time around $t=380$~a.u. (right panel of Fig.~\ref{fig3}). $\eta$ is particularly small at this time, which implies -- based on Eq.~\eqref{eqindic} -- that there is only a small overlap between the two nuclear wavefunction contributions (see Fig.~\ref{fig4}, upper right panel). 

\begin{figure}
\center
\includegraphics[width=1.0\textwidth]{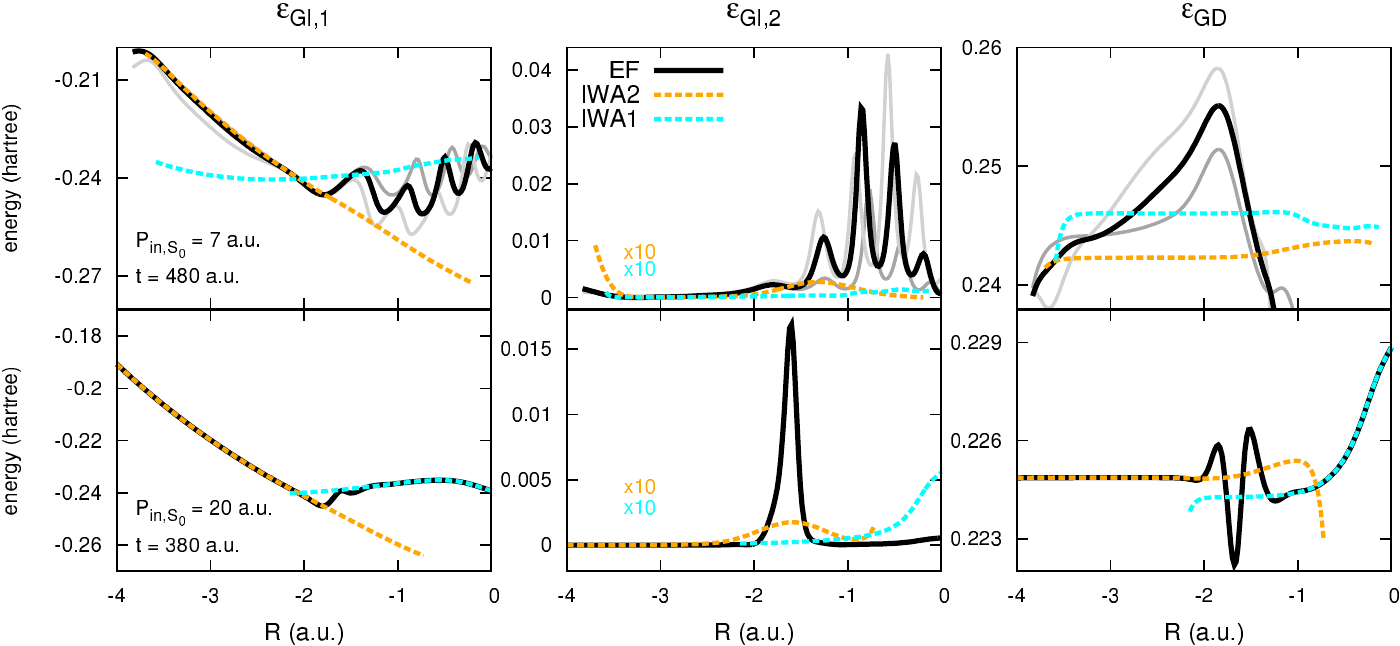}
\caption{Contributions to the full TDPESs represented in Fig.~\ref{fig4}, for the exact (black) and the IWA (cyan and orange) dynamics. Upper panels correspond to the t=480 a.u. in the $P_{in,S_0}= 7$~a.u. run, and the lower panels to t=380 a.u. in the $P_{in,S_0}= 20$~a.u. run. Dashed curves in the last column were rigidly $y$-shifted for clarity. In the upper panels, gray curves show the components of the full TDPESs for $P_{in,S_0}= 6$~a.u. (light gray) and $P_{in,S_0}= 8$~a.u. (dark gray).} 
\label{fig5}
\end{figure}

Comparing the exact TDPES for the full quantum dynamics and the IWA for selected times brings interesting insights (Fig.~\ref{fig4}). Let us start with the case where quantum interferences between the two initial nuclear components are small in the coupling region (right panels of Fig.~\ref{fig4}). For this particular case, the TDPES for the exact dynamics, $\epsilon$, perfectly matches the TDPESs resulting from the two IWA runs, namely $\tilde{\epsilon}_1$ (resulting from the initial conditions given by Eq.~\eqref{iniwf1}) and $\tilde{\epsilon}_2$ (Eq.~\eqref{iniwf2}). The only difference between the exact and IWA dynamics consists in a peak in the region $-2.0<R<-1.0$~a.u. for the exact TDPES. This peak acts as a barrier, preventing a mix between the nuclear wavefunction contributions present on each of its side. This new feature of the exact TDPES resembles the dynamical step observed for single nonadiabatic events~\cite{Gross_PRL2013}. Even though the probability density distribution for the exact nuclear wavefunction perfectly matches the sum of the two individual probability density distributions for the IWA (upper right panel of Fig.~\ref{fig4}), the peak is not formed in the latter case due to the independent character of the IWA dynamics. Upon decomposition of the exact TDPES in its three contributions (Fig.~\ref{fig5}), the origin of this peak appears to come from $\epsilon_{GI,2}$, and is therefore related to the effect of the electron-nuclear coupling operator. The sum of $\epsilon_{GI,1}$ and $\epsilon_{GD}$ would only result in the formation of a small step in the region of the actual peak. As observed in Fig.~\ref{fig5}, the IWA leads to the absence of the barrier in the corresponding $\epsilon_{GI,2}$ term (dashed lines, note that these quantities are multiplied by 10 in the figure).

Such a simple and linear rationalization of the shape of the exact TDPES is unfortunately no more possible whenever important nonadiabatic quantum interferences effects arise, as clearly observed in the left panel of Fig.~\ref{fig4}. In this case, summing the two nuclear probability density distributions for the IWA does not reproduce the exact one, due to the neglect of interferences in the coupling region within the IWA (see also the discussion on Fig.~\ref{fig1}, and Fig.~\ref{fig6}). As a result, the shape of the exact TDPES $\epsilon$ is hardly matched with that of the individual IWA TDPESs $\tilde{\epsilon}_1$ and $\tilde{\epsilon}_2$ (lower left panel of Fig.~\ref{fig4}). The only exception is the portion of configuration space with $R<-3$~a.u., where the shape of $\epsilon$ overlaps with that of $\tilde{\epsilon}_2$ as the nuclear wavefunction is not yet affected by effects in the coupling region. As observed in Fig.~\ref{fig5}, both $\epsilon_{GI,1}$ and $\epsilon_{GI,2}$ match pretty well the IWA1 in this particular region. However, the picture becomes more complicated for the different contributions to the TDPES in and after the coupling region. The two GI parts show strong oscillations, absent from the IWA. The intensity of these oscillations can be related to phase effects, as described with a simple model in App.~\ref{appmodel}. The exact GD term also strongly differs form the corresponding IWA contributions. In fact, the IWA terms have weak structures in $R$-space in the region $-3.0<R<-1.5$~a.u., while the exact GD term has a marked maximum at $R=-1.85$~a.u.. A model for quantum interferences based on Gaussian wavefunctions in presented in the App.~\ref{appmodel}. While this model simplifies the interference picture, it clearly demonstrates the effect of relative phases in the oscillations and the structure of the exact TDPES. In the upper panel of Fig.~\ref{fig5} we show as gray curves the GI and GD components of the TDPES for small variations of the initial momentum around $P_{in,S_0}=7$~a.u.. Even if the GI contributions to TDPES are high oscillatory, the characteristics of the oscillations, i.e., the peaks and their positions, exhibit a smooth dependence on a small variation of the initial condition. The GD part is also smoothly affected by the change of initial condition, mainly in the height of the maximum.

Before concluding this section, let us briefly comment on the correspondence between the BO and EF picture of quantum interferences. In Sec.~\ref{interference} we have identified quantum interferences in the BO framework as fundamentally nuclear quantities, appearing in the expression of the nuclear density~\eqref{eqn: nuclear density with interference}. Moving to the EF perspective, however, it is the TDPES that shows features connected to interferences, and the TDPES depends only on the electronic wavefunction. Indeed, nuclear and electronic dynamics are coupled, meaning that the electronic wavefunction itself depends on the nuclear wavefunction. Still, how can we move the focus of the problem from the BO nuclear wavefunctions to the TDPES? The quantum interference terms in Eq.~\eqref{eqn: nuclear density with interference} can be translated into the EF language as
\begin{align}\label{eqn: interferences from BO to EF}
\left({\chi_{BO}^{(S_0)}}(\bs R,t)\right)^*\overline\chi_{BO}^{(S_0)}(\bs R,t)&+\left({\chi_{BO}^{(S_1)}}(\bs R,t)\right)^*\overline\chi_{BO}^{(S_1)}(\bs R,t)= \notag \\
& \left|\chi(\bs R,t)\right|^2\left[\left(C_{S_0}(\bs R,t)\right)^*\overline C_{S_0}(\bs R,t)+\left(C_{S_1}(\bs R,t)\right)^*\overline C_{S_1}(\bs R,t)\right].
\end{align}
On the right-hand-side, only the nuclear density appears, together with the coefficients of the Born-Huang expansion of the electronic wavefunction $\Phi_{\bs R}(\bs r,t)$ of Eq.~\eqref{eqn: product vs sum}. They are electronic coefficients, even though they depend on $\bs R$, in the sense that, for instance, at a given nuclear position $\bs R$, their squared moduli are the electronic populations for that nuclear configuration. Hence, quantum interferences are mediated only by the electronic coefficients, according to Eq.~\eqref{eqn: interferences from BO to EF}. It is natural, therefore, to expect that such dependence is transmitted to the TDPES.

This section analyzed the behavior of the TDPES when quantum interferences emerge from nonadiabatic events. A simple representation of TDPES based on diabatically-connected BO potential energy surfaces is no more possible in the presence of nonadiabatic quantum interferences, and the TDPES develops new features, such as oscillations, peaks, and mean-field behavior, to name a few.

\subsection{Independent classical trajectories on the exact TDPES}
\label{cltrajs}

As discussed in the last section, the appearance of quantum interferences is transfered from the nuclear wavefunction contributions in the BO representation to the electronic coefficients -- and therefore to the TDPES -- in the EF formalism. This change of perspective from the BO to the EF framework suggests the possibility that Newtonian trajectories evolving on the (exact) TDPES are able to correctly reproduce the quantum nuclear distribution, even in the case studied here where nuclear dynamics manifests a strong quantum-mechanical character related to interference. It is important to underline that the analysis presented below does not propose a strategy to approximate the TDPES. Here, we investigate the potential of classical trajectories to capture nuclear dynamics, provided that a good (in this case, the best) approximation to the TDPES, and thus to the electronic dynamics, is available.

\begin{figure}
\center
\includegraphics[width=1.0\textwidth]{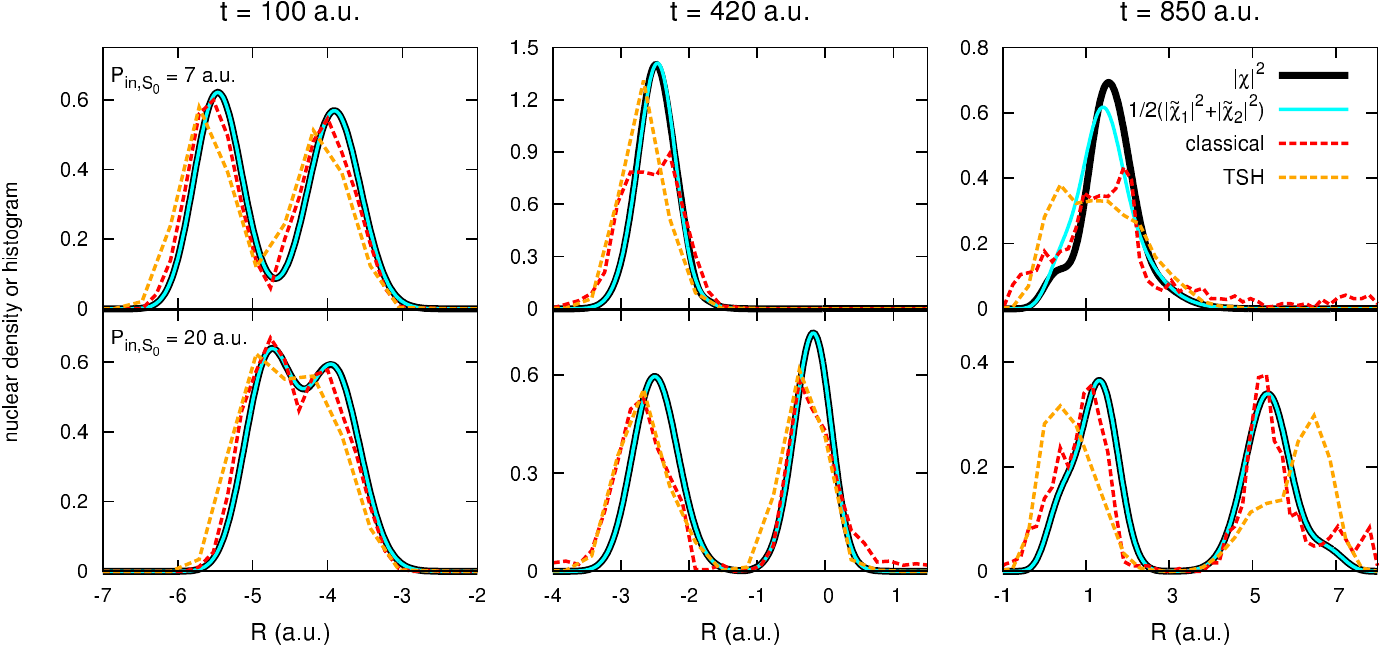}
\caption{Comparison between nuclear probability density distributions (exact and IWA, solid lines) and histograms constructed from the distribution of the classical positions (TSH and classical trajectories evolving on the exact TDPES, dashed lines). Snapshots are given for $t=$100, 420, and 865~a.u., for the runs with $P_{in,S_0}= 7$~a.u. (upper panels) $P_{in,S_0}= 20$~a.u. (lower panels).} 
\label{fig6}
\end{figure}

We have observed that the TDPES exhibits an highly non-trivial dynamical shape when strong interferences takes place in the coupling region. 
Recent works showed that an ensemble of independent classical particles, $|\chi(R,t_0)|^2$-distributed at $t=0$~a.u., closely follows the nuclear probability density distribution at later times when propagated classically on the exact TDPES~\cite{Gross_JCP2015}. Is it still possible to capture the complex nuclear dynamics resulting from interferences in a nonadiabatic region with independent classical trajectories evolving on the TDPES? Fig.~\ref{fig6} answers positively to this question. In the simplest case of weak interferences ($P_{in,S_0}= 20$~a.u., lower panel), the classical trajectories propagated on the exact TDPES nicely reproduce the splitting of the nuclear probability density distribution. In fact, the independent classical trajectories propagated within the EF formalism give a similar, if not slightly better, description of $|\chi(R,t)|^2$ than the TSH independent classical trajectories. As expected from the previous discussions, the probability density from the IWA matches perfectly the $|\chi(R,t)|^2$ distribution for this weakly-interfering case. 
\\
When it comes to stronger interferences ($P_{in,S_0}= 7$~a.u., upper panel of Fig.~\ref{fig6}), the classical independent trajectories -- both in the TSH and EF dynamics -- manage to follow the exact $|\chi(R,t)|^2$. It is interesting to note that, at $t=850$~a.u., the EF classical trajectories distribution exhibits similar features as $|\chi(R,t)|^2$, such as the position of the main peak and the shoulder towards lower R values~\footnote{Numerical convergence of the presented classical simulations on the TDPES was assessed by varying the integration time step in the classical equations of motion, and by refining the spatial grid used to represent the TDPES at every time step.}. For the same time, TSH gives a less structured distribution of trajectories, resulting from a different ratio of trajectories running in each electronic state (Fig.~\ref{fig1}). 
 We note that we only consider here the probability density distribution in R-space, and not the actual projection on the corresponding BO states. As mentioned before, the IWA probability density distribution does not exactly follow $|\chi(R,t)|^2$, whenever interferences start to play an important role ($t>100$~a.u.).

\section{Conclusions}
\label{conclusion}
In this work, we have analyzed the problem of quantum interferences in nonadiabatic dynamics from the exact factorization perspective. In this context, we have addressed two main questions {\color{black}by studying a one-dimensional model of nonadiabatic electron transfer, where characteristic features of quantum interferences already appear}. The first is whether, and how, the appearance of interferences can be identified in the TDPES, the potential that drives the nuclear evolution. The second is whether classical (independent) trajectories provide a correct description of nuclear motion when electronic dynamics, and thus the TDPES, is solved exactly. In the spirit of previous work~\cite{Gross_JCP2015} that has led to the derivation of a trajectory-based quantum-classical algorithm~\cite{Gross_PRL2015, Gross_JCTC2016} to describe nonadiabatic processes, the aim of our study is to assess the potential of a similar treatment of a problem that now present a profound quantum-mechanical character, i.e., quantum interferences. We stress that this work should be regarded as an exploratory study, rather than as the development of an actual numerical procedure.
{\color{black}It should be borne in mind that quantum-classical approaches are powerful tools to investigate the coupled dynamics of electrons and nuclei in complex (and large) molecular systems -- for which a fully quantum-mechanical description is prohibitively expensive. However, introducing a classical approximation for selected nuclear degrees of freedom might come at the price of missing some features of the dynamics such as nonadiabatic quantum interferences, whose importance has not been considered in previous work on the exact factorization. 
The analysis proposed here identifies such quantum features in the TDPES, and it furthermore generalizes the use of a quantum-classical dynamics based on the exact factorization for nonadiabatic quantum interferences.}

{\color{black}In answering the first question above, related to the appearance of interferences in the TDPES,} we have pointed out the importance of focusing the analysis in the nonadiabatic coupling region. In this region, BO wavefunctions with different histories recombine and effects connected to their relative phases become evident in the TDPES. We have observed, in fact, the appearance of an oscillatory behavior of the TDPES that initiates during the nonadiabatic event. We have rationalized such behavior with a simplified model that allowed us to control the phase relation between the incoming BO wavefunctions and to tune the oscillations in the TDPES.

Surprisingly, the answer to the second question is that classical trajectories evolved on the exact TDPES are able to reproduce {\color{black} rather well} the quantum probability distribution at all times even in the case when quantum interferences appear. This test proves once again that the TDPES is a powerful tool to drive nuclear dynamics in nonadiabatic conditions, and thus confirms the importance of developing accurate approximation schemes to compute it in general cases.

More generally, this study opens interesting questions related to the BO and the exact factorization framework. The question often arose in the literature as to whether or not Newtonian (classical) trajectories are an efficient tool to mimic quantum nuclear dynamics in nonadiabatic conditions (see Refs.~\cite{deumens1994time,Lopreore2002,PhysRevA.71.032511,takatsuka2006non,irenebook,shenvi2009phase,curchod2013ontrajectory,malhado2014non,Gross_JCTC2016} for examples). From a BO perspective, nonadiabatic regions are the heart of the problem, due to the fact that nonadiabatic couplings are present in the (a priori infinite) coupled evolution equations for the BO wavefunctions. 
In this framework, we might argue that such involved nuclear evolution equations compensate the rather simple time-independent electronic problem. When the same situation is analyzed from the exact factorization perspective, the complexity of the coupled electron-nuclear problem is somehow shifted towards the electronic problem and towards the calculation of the TDPES (and, when necessary, of the time-dependent vector potential). At the cost of introducing an actual time-dependence in the electronic wavefunction, the nuclear equation becomes a standard time-dependent Schr\"odinger equation, which can be approximated in terms of Newtonian trajectories. Providing accurate approximations to the TDPES might nevertheless be a hard task, and we shall resort in compromising between the simplification of the nuclear dynamics and the solution of the electronic problem (see for instance the idea of using coupled trajectories~\cite{Gross_PRL2015,Gross_JCTC2016} to solve the nuclear equation of the exact factorization). Still, developing methods that play with distributing the electron-nuclear coupling complexity between the electronic and the nuclear problem in the context of the exact factorization is an open subject, reinforced by a better understanding of the exact factorization quantities as done in the present work.

\appendix

\section{A simple model of nonadiabatic quantum interferences between nuclear wavefunctions}
\label{appmodel}
In some of the situations studied above, the TDPES presents a more complex structure than previous observations~\cite{Gross_PRL2013} have pointed out and interpreted. When quantum interferences (as defined in Sec.~\ref{interference}) are observed, oscillations in the gauge-invariant components of the TDPES seem to appear (in Fig.~\ref{fig5} for the case $P_{in,S_0}=7$~a.u.) at and after the passage of the nuclear wavefunction through the coupling region. The aim of this Appendix is to rationalize the appearance of these oscillations and to relate them to the relative phases of different BO nuclear wavefunctions, suggesting in this way a connection between the BO and the EF frameworks. To this end, we construct a model that allows us to control the relative phases and to directly relate them to changes in the (exact) nuclear density and in the TDPES. Comparisons with the IWA applied to the model will be also discussed. Below, we will first introduce the model, and we will then discuss the properties of the TDPES.

\begin{figure}
\center
\includegraphics[width=.50\textwidth]{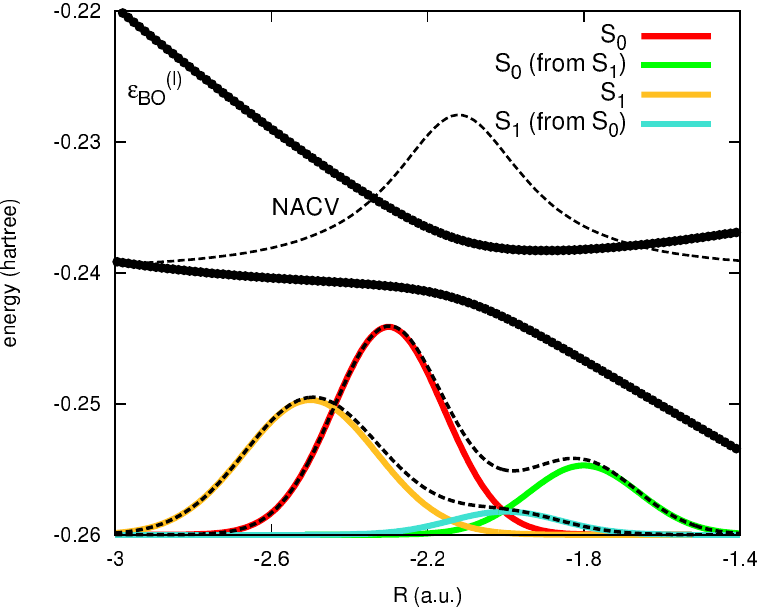}
\caption{Model used to analyze the nuclear density and TDPES at the avoided crossing. The thick black lines are the adiabatic potential energy curves, $\epsilon_{BO}^{(l)}(R)$, while the dashed thin line (NACV) shows the quantity $\langle\Phi_{R}^{(1)}|\partial_R\Phi_{R}^{(2)}\rangle_r$. The Gaussian densities corresponding to $S_0$ are plotted as red and green lines; the densities corresponding to $S_1$ are plotted as orange and cyan lines. The corresponding density envelops from Eqs.~\eqref{eqn: model wp S0} and~\eqref{eqn: model wp S1} for the value $dP=0$~a.u. are shown as dashed black lines.}
\label{fig7}
\end{figure}

When the nuclear wavepacket $\chi_{BO}^{(S_0)}(R,t)$, initially evolving along the $S_0$ surface, crosses the coupling region, it transfers amplitude to $S_1$; similarly, $\chi_{BO}^{(S_1)}(R,t)$ transfers amplitude to $S_0$ (see Fig.~\ref{fig0bis}). In the model, we suppose that at a certain time $t^*$ during the transfer, $\chi_{BO}^{(S_0)}(R,t^*)$ can be approximated as the sum of two contributions, one associated to the wavepacket on $S_0$ that has not yet reached the avoided crossing, and the other associated to the wavepacket created on $S_0$ after the transfer from $S_1$ (see Eq.~\eqref{eqn: Psi at time t''} for comparison). Moreover, these two contributions are Gaussian wavepackets travelling with momentum $P+dP$ and $P$, respectively, and are centered at different positions. Therefore, the expression adopted in the model for $\chi_{BO}^{(S_0)}(R,t^*)$ is
\begin{align}\label{eqn: model wp S0}
\chi_{BO}^{(S_0)}(R,t^*) = N_0
\left[G_{\sigma_0}\left(R-R_0\right)e^{\frac{i}{\hbar}(P+dP)(R-R_{0})}+
G_{\sigma_{0\leftarrow1}}\left(R-R_{0\leftarrow1}\right)e^{\frac{i}{\hbar}P(R-R_{0\leftarrow1})}\right],
\end{align}
where the subscript $0$ label the quantities associated to the wavepacket evolving (and remaining) on $S_0$, while the subscript $0\leftarrow1$ corresponds to the wavepacket transferred from $S_1$. In Fig.~\ref{fig7}, the squared modulus of the first contribution of Eq.~\eqref{eqn: model wp S0} is shown in red, while the second contribution is shown in green. The same idea applies to $\chi_{BO}^{(S_1)}(R,t^*)$, during the amplitude transfer from $S_1$ to $S_0$. The (squared moduli of the) two contributions are shown in Fig.~\ref{fig7}, with an orange line for that centered at $R_1$ with variance $\sigma_{1}$ and associated to a momentum $P$, and a cyan line for the contribution centered at $R_{1\leftarrow0}$ with variance $\sigma_{1\leftarrow0}$ associated to a momentum $P+dP$. The expression of $\chi_{BO}^{(S_1)}(R,t^*)$ is
\begin{align}\label{eqn: model wp S1}
\chi_{BO}^{(S_1)}(R,t^*) = N_1 \left[G_{\sigma_1}\left(R-R_1\right)e^{\frac{i}{\hbar}P(R-R_{1})}+G_{\sigma_{1\leftarrow0}}\left(R-R_{1\leftarrow0}\right)e^{\frac{i}{\hbar}(P+dP)(R-R_{1\leftarrow0})}\right].
\end{align}
The parameters used in the numerical calculations are the following: the normalization constants in the expressions of the nuclear wavefunctions are chosen such that $\int dR|\chi_{BO}^{(S_0)}(R,t^*)|^2=\int dR\chi_{BO}^{(S_1)}(R,t^*)=0.5$; the centers of the Gaussians are $R_{0}=-2.3$~a.u., $R_{0\leftarrow1}=-1.8$~a.u., $R_{1}=-2.5$~a.u. and $R_{1\leftarrow0}=-2.0$~a.u.; the variances of the Gaussians are $\sigma_{0}=0.20$~a.u., $\sigma_{0\leftarrow1}=0.20$~a.u. and $\sigma_{1}=0.20$~a.u., $\sigma_{1\leftarrow0}=0.24$~a.u.; the momentum $P$ is fixed to the value $3$~a.u., while $dP$ is varied in the range 0 to 10~a.u.. The idea here is that the wavepacket on one BO curve approaching the coupling region with a certain momentum, i.e., either $P$ or $P+dP$, produces on the other BO curve a contribution, which initially has the same momentum. Furthermore, since all above calculations have been performed by fixing the momentum of $\chi_{BO}^{(S_1)}(R,t_0)$ to 0~a.u. and by varying $P_{in,S_0}$ in the range -10~a.u. to 20~a.u., we assume here that $\chi_{BO}^{(S_1)}(R,t^*)$ has acquired momentum when it reaches the avoided crossing, we fix this value, and we vary only the momentum of $\chi_{BO}^{(S_0)}(R,t^*)$ by varying $dP$.

Using this form for the BO nuclear wavefunctions at the time of the nonadiabatic event, we compute the full molecular wavefunction from Eq.~\eqref{iniwf}, the nuclear wavefunction using the gauge condition derived from Eq.~\eqref{fixing the gauge}, and finally the TDPES. In particular, Gaussian-shaped wavefunctions with well-defined momenta allow us (i) to obtain smooth functions of $R$ when spatial derivatives are involved, as in the expression of $\epsilon_{GI,2}$, and (ii) to directly associate the changes in the TDPES to the variation of $dP$. Since a real time-evolution is not simulated, we do not have direct access to the gauge-dependent part of the TDPES, which contains the time-derivative of the electronic wavefunction. However, following the analysis of Ref.~\cite{Gross_JCP2015}, we approximate $\epsilon_{GD}(R,t^*)$ as
\begin{align}\label{eqn: approximation for eGD}
\epsilon_{GD}(R,t^*) \simeq \int^R dR' \left(\epsilon_{BO}^{(S_1)}(R')-\epsilon_{BO}^{(S_1)}(R')\right) \partial_{R'} \left|C_{S_0}(R',t^*)\right|^2.
\end{align}
While this approximate expression is valid in regions where the nonadiabatic couplings are negligible -- which is indeed not the case considered here -- this contribution has been shown to be the leading one in a case of single nonadiabatic event. It will furthermore give us  an idea of the overall shape of $\epsilon_{GD}(R,t^*)$.

\begin{figure}
\center
\includegraphics[width=.70\textwidth]{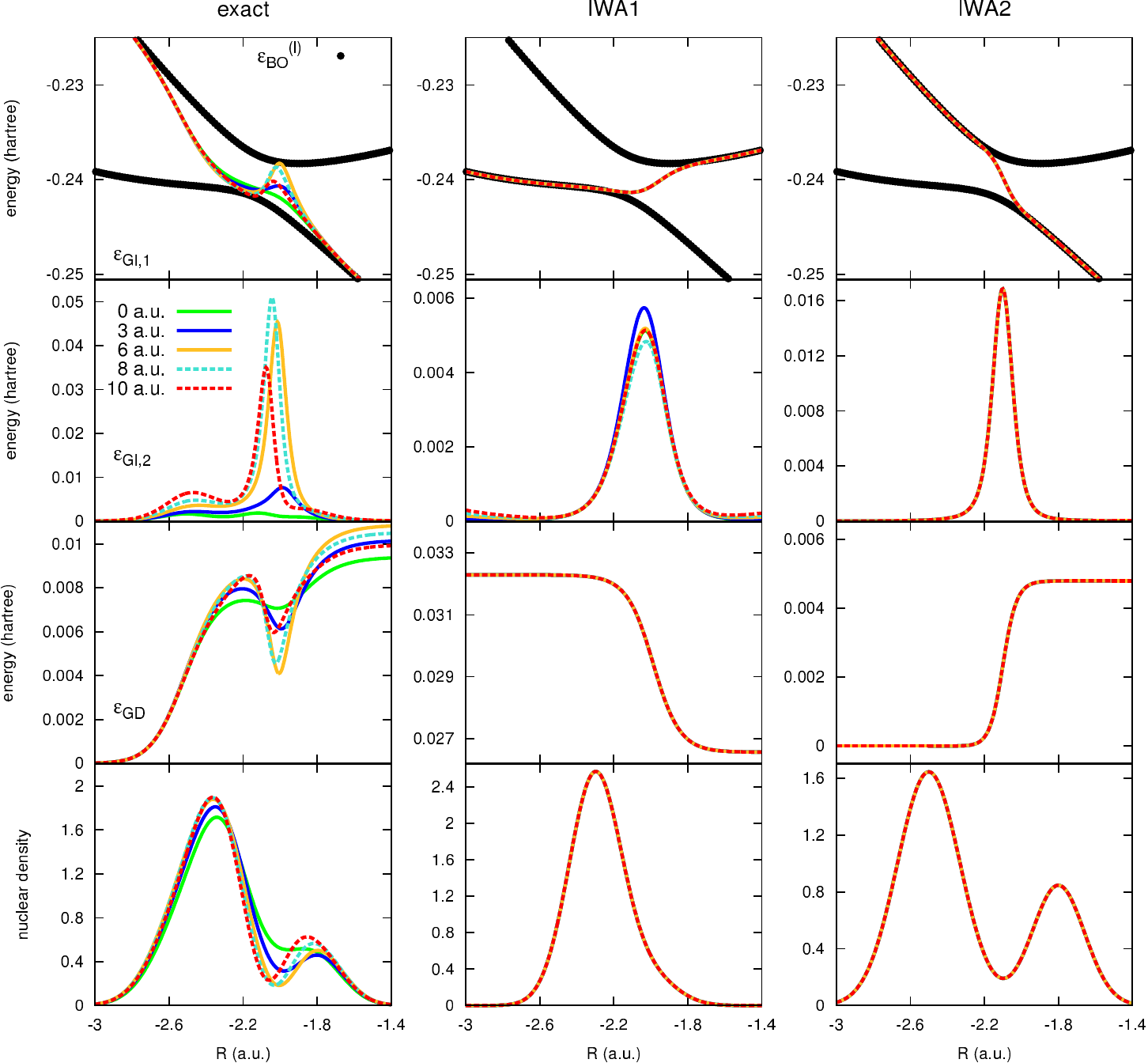}
\caption{TDPES and nuclear densities for the model presented in Fig.~\ref{fig7}. Left panels refer to exact calculations, while the central and right panels refer to the IWA, as described in the text. From top to bottom the plots show $\epsilon_{GI,1}(R,t^*)$ in colors and the BO curves (black dotted curves), $\epsilon_{GI,2}(R,t^*)$, $\epsilon_{GD}(R,t^*)$ from Eq.~\eqref{eqn: approximation for eGD}, and $|\chi(R,t^*)|^2$. The colors refer to different values of $dP$ as shown in the plots.} 
\label{fig8}
\end{figure}

Fig.~\ref{fig8} shows the numerical results for different values of $dP$. The panels on the left present the three components of the TDPES, $\epsilon_{GI,1}(R,t^*)$ (top), $\epsilon_{GI,2}(R,t^*)$ (second from the top) and $\epsilon_{GD}(R,t^*)$ (third from the top), and the nuclear density (bottom). The colors correspond to some selected values of $dP$ as indicated in the figure. In the regions corresponding to the tails of the nuclear densities, $\epsilon_{GI,1}(R,t^*)$ follows, as expected, either one or the other BO potential energy curve independently of $dP$. In the intermediate region small oscillations are observed: the oscillations occur within the limits represented by the adiabatic curves and have different amplitudes depending on the value of $dP$. The amplitudes, however, do not have a monotonic behavior as functions of $dP$, as they increase up to $dP=6$~a.u. and they decrease for larger values. A similar structure is observed in the shape of $\epsilon_{GI,2}(R,t^*)$, where also the displacement of the maxima is clearly observed as effect of changes in $dP$. The gauge-dependent part of the TDPES does not present a simple step-like shape, as observed in the case of a single crossing event~\cite{Gross_PRL2013}, but oscillations appear, mirroring the oscillations in $\epsilon_{GI,1}(R,t^*)$. At the tails, both $\epsilon_{GI,2}(R,t^*)$ and $\epsilon_{GD}(R,t^*)$ reach constant values, not altering the slope of $\epsilon_{GI,1}(R,t^*)$. It is interesting to notice that the peaks in $\epsilon_{GI,2}(R,t^*)$ might be larger than those in the other two components of the TDPES, thus they drastically affect the shape of the potential in the intermediate region and, consequently, strongly affect the nuclear dynamics: the passage of the nuclear wavefunction through the coupling region, and thus the ``amount'' of nuclear density transferred through the crossing, is controlled by the presence of the potential barrier in $\epsilon_{GI,2}(R,t^*)$. 

In this model we cannot reproduce all the oscillations observed in Fig.~\ref{fig5} since in the tail regions the Gaussian wavefunctions go monotonically to zero, while in the full dynamics the BO wavefunctions present additional oscillations in these regions.

The shape of the TDPES in the nonadiabatic region seems then to be strongly dependent on the phase relation between the BO wavefunctions approaching the coupling region. In order to further validate this statement, we analyze the results of the IWA for both states $S_0$ and $S_1$. In this case, we compute the three terms of the TDPES by using the expressions 
\begin{align}
\tilde\Psi_1(r,R,t^*)= \tilde\chi_{1}^{(S_0)}(R,t^*) \Phi_{BO}^{(S_0)}(r)+\tilde\chi_{1}^{(S_1)}(R,t^*) \Phi_{BO}^{(S_1)}(r)\label{eqn: iwa1 in model}
\end{align}
and
\begin{align}
\tilde\Psi_2(r,R,t^*)= \tilde\chi_{2}^{(S_0)}(R,t^*) \Phi_{BO}^{(S_0)}(r)+\tilde\chi_{2}^{(S_1)}(R,t^*) \Phi_{BO}^{(S_1)}(r).\label{eqn: iwa2 in model}
\end{align}
Here, $\tilde\chi_{1}^{(S_0)}(R,t^*)$ contains only the first term on the right-hand-side of Eq.~\eqref{eqn: model wp S0} while $\tilde\chi_{1}^{(S_1)}(R,t^*)$ contains only the second term on the right-hand-side of Eq.~\eqref{eqn: model wp S1}. Vice versa, $\tilde\chi_{2}^{(S_0)}(R,t^*)$ and $\tilde\chi_{2}^{(S_1)}(R,t^*)$ contain the second term on the right-hand-side of Eq.~\eqref{eqn: model wp S0} and the first term on the right-hand-side of Eq.~\eqref{eqn: model wp S1}, respectively. This choice aims at reproducing the idea behind the IWA used to analyze the appearance and effect of interferences on the full dynamics discussed in Sec.~\ref{results}.

In both expressions~(\ref{eqn: iwa1 in model}) and~(\ref{eqn: iwa2 in model}), the BO wavefunctions have the same phase, which indeed disappears in the components of the TDPES related to the squared moduli of $C_{S_0}(R,t^*)$ or $C_{S_1}(R,t^*)$ and, consequently, to the squared moduli of the BO nuclear wavefunctions. Therefore, it is easy to understand the structure of the TDPES and of the nuclear density in Fig.~\ref{fig8} (central and right panels). All curves, for different values of $dP$, have the same behavior. Only some terms in the expression of $\epsilon_{GI,2}$ do not simply depend on squared moduli of $C_{S_0}(R,t^*)$ or $C_{S_1}(R,t^*)$ and this is the reason why some curves do not exactly have the same shape. The IWA calculations do not show multiple peaks in the components of the TDPES, as they have the simple shapes already observed~\cite{Gross_PRL2013} in the case of a single nonadiabatic crossing event. Interference effects are completely absent in the IWA, thus we can clearly ascribe the appearance of the oscillatory structure of the TDPES to the relative phases of the BO nuclear wavefunctions.


\begin{thebibliography}{95}
\expandafter\ifx\csname natexlab\endcsname\relax\def\natexlab#1{#1}\fi
\expandafter\ifx\csname bibnamefont\endcsname\relax
  \def\bibnamefont#1{#1}\fi
\expandafter\ifx\csname bibfnamefont\endcsname\relax
  \def\bibfnamefont#1{#1}\fi
\expandafter\ifx\csname citenamefont\endcsname\relax
  \def\citenamefont#1{#1}\fi
\expandafter\ifx\csname url\endcsname\relax
  \def\url#1{\texttt{#1}}\fi
\expandafter\ifx\csname urlprefix\endcsname\relax\def\urlprefix{URL }\fi
\providecommand{\bibinfo}[2]{#2}
\providecommand{\eprint}[2][]{\url{#2}}

\bibitem[{\citenamefont{Born and Oppenheimer}(1927)}]{born1927quantentheorie}
\bibinfo{author}{\bibfnamefont{M.}~\bibnamefont{Born}} \bibnamefont{and}
  \bibinfo{author}{\bibfnamefont{R.}~\bibnamefont{Oppenheimer}},
  \bibinfo{journal}{Annalen der Physik} \textbf{\bibinfo{volume}{389}},
  \bibinfo{pages}{457} (\bibinfo{year}{1927}).

\bibitem[{\citenamefont{Tully}(2000)}]{tully2000perspective}
\bibinfo{author}{\bibfnamefont{J.~C.} \bibnamefont{Tully}},
  \bibinfo{journal}{Theor. Chem. Acc.} \textbf{\bibinfo{volume}{103}},
  \bibinfo{pages}{173} (\bibinfo{year}{2000}).

\bibitem[{\citenamefont{Beck et~al.}(2000)\citenamefont{Beck, J\"{a}ckle,
  Worth, and Meyer}}]{Beck2000}
\bibinfo{author}{\bibfnamefont{M.~H.} \bibnamefont{Beck}},
  \bibinfo{author}{\bibfnamefont{A.}~\bibnamefont{J\"{a}ckle}},
  \bibinfo{author}{\bibfnamefont{G.~A.} \bibnamefont{Worth}}, \bibnamefont{and}
  \bibinfo{author}{\bibfnamefont{H.~D.} \bibnamefont{Meyer}},
  \bibinfo{journal}{Phys. Rep.} \textbf{\bibinfo{volume}{324}},
  \bibinfo{pages}{1} (\bibinfo{year}{2000}).

\bibitem[{\citenamefont{Worth et~al.}(2004)\citenamefont{Worth, Meyer, and
  Cederbaum}}]{worth04}
\bibinfo{author}{\bibfnamefont{G.~A.} \bibnamefont{Worth}},
  \bibinfo{author}{\bibfnamefont{H.~D.} \bibnamefont{Meyer}}, \bibnamefont{and}
  \bibinfo{author}{\bibfnamefont{L.~S.} \bibnamefont{Cederbaum}}, in
  \emph{\bibinfo{booktitle}{Conical Intersections: Electronic Structure,
  Dynamics \& Spectroscopy}}, edited by
  \bibinfo{editor}{\bibfnamefont{W.}~\bibnamefont{Domcke}},
  \bibinfo{editor}{\bibfnamefont{D.~R.} \bibnamefont{Yarkony}},
  \bibnamefont{and} \bibinfo{editor}{\bibfnamefont{H.}~\bibnamefont{K\"oppel}}
  (\bibinfo{year}{2004}), vol.~\bibinfo{volume}{15} of
  \emph{\bibinfo{series}{Advanced Series in Physical Chemistry}},
  chap.~\bibinfo{chapter}{14}, pp. \bibinfo{pages}{583--617}.

\bibitem[{\citenamefont{Persico and Granucci}(2014)}]{persico2014overview}
\bibinfo{author}{\bibfnamefont{M.}~\bibnamefont{Persico}} \bibnamefont{and}
  \bibinfo{author}{\bibfnamefont{G.}~\bibnamefont{Granucci}},
  \bibinfo{journal}{Theor. Chem. Acc.} \textbf{\bibinfo{volume}{133}},
  \bibinfo{pages}{1} (\bibinfo{year}{2014}).

\bibitem[{\citenamefont{Tully and Preston}(1971)}]{preston71}
\bibinfo{author}{\bibfnamefont{J.~C.} \bibnamefont{Tully}} \bibnamefont{and}
  \bibinfo{author}{\bibfnamefont{R.~K.} \bibnamefont{Preston}},
  \bibinfo{journal}{J. Chem. Phys.} \textbf{\bibinfo{volume}{55}},
  \bibinfo{pages}{562} (\bibinfo{year}{1971}).

\bibitem[{\citenamefont{Tully}(1990)}]{tully90}
\bibinfo{author}{\bibfnamefont{J.~C.} \bibnamefont{Tully}},
  \bibinfo{journal}{J. Chem. Phys.} \textbf{\bibinfo{volume}{93}},
  \bibinfo{pages}{1061} (\bibinfo{year}{1990}).

\bibitem[{\citenamefont{Coker and Xiao}(1995)}]{coker1995methods}
\bibinfo{author}{\bibfnamefont{D.}~\bibnamefont{Coker}} \bibnamefont{and}
  \bibinfo{author}{\bibfnamefont{L.}~\bibnamefont{Xiao}}, \bibinfo{journal}{J.
  Chem. Phys.} \textbf{\bibinfo{volume}{102}}, \bibinfo{pages}{496}
  (\bibinfo{year}{1995}).

\bibitem[{\citenamefont{Bittner and Rossky}(1995)}]{Bittner1995}
\bibinfo{author}{\bibfnamefont{E.~R.} \bibnamefont{Bittner}} \bibnamefont{and}
  \bibinfo{author}{\bibfnamefont{P.~J.} \bibnamefont{Rossky}},
  \bibinfo{journal}{J. Chem. Phys.} \textbf{\bibinfo{volume}{103}},
  \bibinfo{pages}{8130} (\bibinfo{year}{1995}).

\bibitem[{\citenamefont{Prezhdo and Rossky}(1997)}]{prezhdo97}
\bibinfo{author}{\bibfnamefont{O.~V.} \bibnamefont{Prezhdo}} \bibnamefont{and}
  \bibinfo{author}{\bibfnamefont{P.~J.} \bibnamefont{Rossky}},
  \bibinfo{journal}{J. Chem. Phys.} \textbf{\bibinfo{volume}{107}},
  \bibinfo{pages}{825} (\bibinfo{year}{1997}).

\bibitem[{\citenamefont{Tully}(1998)}]{Tully1998}
\bibinfo{author}{\bibfnamefont{J.~C.} \bibnamefont{Tully}},
  \bibinfo{journal}{Faraday Discuss.} \textbf{\bibinfo{volume}{110}},
  \bibinfo{pages}{407} (\bibinfo{year}{1998}).

\bibitem[{\citenamefont{Kapral and Ciccotti}(1999)}]{kapral:8919}
\bibinfo{author}{\bibfnamefont{R.}~\bibnamefont{Kapral}} \bibnamefont{and}
  \bibinfo{author}{\bibfnamefont{G.}~\bibnamefont{Ciccotti}},
  \bibinfo{journal}{J. Chem. Phys.} \textbf{\bibinfo{volume}{110}},
  \bibinfo{pages}{8919} (\bibinfo{year}{1999}).

\bibitem[{\citenamefont{Burghardt and
  Cederbaum}(2001)}]{burghardt2001hydrodynamic}
\bibinfo{author}{\bibfnamefont{I.}~\bibnamefont{Burghardt}} \bibnamefont{and}
  \bibinfo{author}{\bibfnamefont{L.}~\bibnamefont{Cederbaum}},
  \bibinfo{journal}{J. Chem. Phys.} \textbf{\bibinfo{volume}{115}},
  \bibinfo{pages}{10312} (\bibinfo{year}{2001}).

\bibitem[{\citenamefont{Wyatt et~al.}(2001)\citenamefont{Wyatt, Lopreore, and
  Parlant}}]{Wyatt2001a}
\bibinfo{author}{\bibfnamefont{R.~E.} \bibnamefont{Wyatt}},
  \bibinfo{author}{\bibfnamefont{C.~L.} \bibnamefont{Lopreore}},
  \bibnamefont{and} \bibinfo{author}{\bibfnamefont{G.}~\bibnamefont{Parlant}},
  \bibinfo{journal}{J. Chem. Phys.} \textbf{\bibinfo{volume}{114}},
  \bibinfo{pages}{5113} (\bibinfo{year}{2001}).

\bibitem[{\citenamefont{Marx and Hutter}(2009)}]{marxbook}
\bibinfo{author}{\bibfnamefont{D.}~\bibnamefont{Marx}} \bibnamefont{and}
  \bibinfo{author}{\bibfnamefont{J.}~\bibnamefont{Hutter}},
  \emph{\bibinfo{title}{Ab Initio Molecular Dynamics: Basic Theory and Advanced
  Methods}} (\bibinfo{publisher}{Cambridge University Press},
  \bibinfo{year}{2009}).

\bibitem[{\citenamefont{Horenko et~al.}(2002)\citenamefont{Horenko, Salzmann,
  Schmidt, and Schutte}}]{Horenko2002}
\bibinfo{author}{\bibfnamefont{I.}~\bibnamefont{Horenko}},
  \bibinfo{author}{\bibfnamefont{C.}~\bibnamefont{Salzmann}},
  \bibinfo{author}{\bibfnamefont{B.}~\bibnamefont{Schmidt}}, \bibnamefont{and}
  \bibinfo{author}{\bibfnamefont{C.}~\bibnamefont{Schutte}},
  \bibinfo{journal}{J. Chem. Phys.} \textbf{\bibinfo{volume}{117}},
  \bibinfo{pages}{11075} (\bibinfo{year}{2002}).

\bibitem[{\citenamefont{Kapral}(2006)}]{doi:10.1146/annurev.physchem.57.032905.104702}
\bibinfo{author}{\bibfnamefont{R.}~\bibnamefont{Kapral}},
  \bibinfo{journal}{Annu. Rev. Phys. Chem.} \textbf{\bibinfo{volume}{57}},
  \bibinfo{pages}{129} (\bibinfo{year}{2006}).

\bibitem[{\citenamefont{Jasper et~al.}(2006)\citenamefont{Jasper, Nangia, Zhu,
  and Truhlar}}]{Jasper2006}
\bibinfo{author}{\bibfnamefont{A.~W.} \bibnamefont{Jasper}},
  \bibinfo{author}{\bibfnamefont{S.}~\bibnamefont{Nangia}},
  \bibinfo{author}{\bibfnamefont{C.}~\bibnamefont{Zhu}}, \bibnamefont{and}
  \bibinfo{author}{\bibfnamefont{D.~G.} \bibnamefont{Truhlar}},
  \bibinfo{journal}{Acc. Chem. Res.} \textbf{\bibinfo{volume}{39}},
  \bibinfo{pages}{101} (\bibinfo{year}{2006}).

\bibitem[{\citenamefont{Barbatti}(2011)}]{barbatti2011}
\bibinfo{author}{\bibfnamefont{M.}~\bibnamefont{Barbatti}},
  \bibinfo{journal}{WIREs Comput. Mol. Sci.} \textbf{\bibinfo{volume}{1}},
  \bibinfo{pages}{620} (\bibinfo{year}{2011}), ISSN \bibinfo{issn}{1759-0884}.

\bibitem[{\citenamefont{Zamstein and
  Tannor}(2012{\natexlab{a}})}]{zamstein2012non}
\bibinfo{author}{\bibfnamefont{N.}~\bibnamefont{Zamstein}} \bibnamefont{and}
  \bibinfo{author}{\bibfnamefont{D.~J.} \bibnamefont{Tannor}},
  \bibinfo{journal}{J. Chem. Phys.} \textbf{\bibinfo{volume}{137}},
  \bibinfo{pages}{22A517} (\bibinfo{year}{2012}{\natexlab{a}}).

\bibitem[{\citenamefont{Zamstein and Tannor}(2012{\natexlab{b}})}]{zamstein2}
\bibinfo{author}{\bibfnamefont{N.}~\bibnamefont{Zamstein}} \bibnamefont{and}
  \bibinfo{author}{\bibfnamefont{D.~J.} \bibnamefont{Tannor}},
  \bibinfo{journal}{J. Chem. Phys.} \textbf{\bibinfo{volume}{137}},
  \bibinfo{pages}{22A518} (\bibinfo{year}{2012}{\natexlab{b}}).

\bibitem[{\citenamefont{Gorshkov et~al.}(2013)\citenamefont{Gorshkov, Tretiak,
  and Mozyrsky}}]{Gorshkov2013}
\bibinfo{author}{\bibfnamefont{V.~N.} \bibnamefont{Gorshkov}},
  \bibinfo{author}{\bibfnamefont{S.}~\bibnamefont{Tretiak}}, \bibnamefont{and}
  \bibinfo{author}{\bibfnamefont{D.}~\bibnamefont{Mozyrsky}},
  \bibinfo{journal}{Nat. Commun.} \textbf{\bibinfo{volume}{4}},
  \bibinfo{pages}{2144} (\bibinfo{year}{2013}).

\bibitem[{\citenamefont{Curchod and
  Tavernelli}(2013)}]{curchod2013ontrajectory}
\bibinfo{author}{\bibfnamefont{B.~F.~E.} \bibnamefont{Curchod}}
  \bibnamefont{and}
  \bibinfo{author}{\bibfnamefont{I.}~\bibnamefont{Tavernelli}},
  \bibinfo{journal}{J. Chem. Phys.} \textbf{\bibinfo{volume}{138}},
  \bibinfo{pages}{184112} (\bibinfo{year}{2013}).

\bibitem[{\citenamefont{Albareda et~al.}(2014)\citenamefont{Albareda, Appel,
  Franco, Abedi, and Rubio}}]{albareda2014correlated}
\bibinfo{author}{\bibfnamefont{G.}~\bibnamefont{Albareda}},
  \bibinfo{author}{\bibfnamefont{H.}~\bibnamefont{Appel}},
  \bibinfo{author}{\bibfnamefont{I.}~\bibnamefont{Franco}},
  \bibinfo{author}{\bibfnamefont{A.}~\bibnamefont{Abedi}}, \bibnamefont{and}
  \bibinfo{author}{\bibfnamefont{A.}~\bibnamefont{Rubio}},
  \bibinfo{journal}{Phys. Rev. Lett.} \textbf{\bibinfo{volume}{113}},
  \bibinfo{pages}{083003} (\bibinfo{year}{2014}).

\bibitem[{\citenamefont{Albareda et~al.}(2015)\citenamefont{Albareda, Bofill,
  Tavernelli, Huarte-Larra{\~n}aga, Illas, and Rubio}}]{guilleandivano}
\bibinfo{author}{\bibfnamefont{G.}~\bibnamefont{Albareda}},
  \bibinfo{author}{\bibfnamefont{J.~M.} \bibnamefont{Bofill}},
  \bibinfo{author}{\bibfnamefont{I.}~\bibnamefont{Tavernelli}},
  \bibinfo{author}{\bibfnamefont{F.}~\bibnamefont{Huarte-Larra{\~n}aga}},
  \bibinfo{author}{\bibfnamefont{F.}~\bibnamefont{Illas}}, \bibnamefont{and}
  \bibinfo{author}{\bibfnamefont{A.}~\bibnamefont{Rubio}}, \bibinfo{journal}{J.
  Chem. Phys. Lett.} \textbf{\bibinfo{volume}{6}}, \bibinfo{pages}{1529}
  (\bibinfo{year}{2015}), \bibinfo{note}{pMID: 26263307}.

\bibitem[{\citenamefont{Mart\'{\i}nez et~al.}(1996)\citenamefont{Mart\'{\i}nez,
  Ben-Nun, and Levine}}]{martinez1996multi}
\bibinfo{author}{\bibfnamefont{T.~J.} \bibnamefont{Mart\'{\i}nez}},
  \bibinfo{author}{\bibfnamefont{M.}~\bibnamefont{Ben-Nun}}, \bibnamefont{and}
  \bibinfo{author}{\bibfnamefont{R.~D.} \bibnamefont{Levine}},
  \bibinfo{journal}{J. Phys. Chem.} \textbf{\bibinfo{volume}{100}},
  \bibinfo{pages}{7884} (\bibinfo{year}{1996}).

\bibitem[{\citenamefont{Ben-Nun and Mart\'{\i}nez}(1998)}]{Ben-Nun1998}
\bibinfo{author}{\bibfnamefont{M.}~\bibnamefont{Ben-Nun}} \bibnamefont{and}
  \bibinfo{author}{\bibfnamefont{T.~J.} \bibnamefont{Mart\'{\i}nez}},
  \bibinfo{journal}{J. Chem. Phys.} \textbf{\bibinfo{volume}{108}},
  \bibinfo{pages}{7244} (\bibinfo{year}{1998}).

\bibitem[{\citenamefont{Ben-Nun and Mart\'{\i}nez}(2002)}]{toddaims}
\bibinfo{author}{\bibfnamefont{M.}~\bibnamefont{Ben-Nun}} \bibnamefont{and}
  \bibinfo{author}{\bibfnamefont{T.~J.} \bibnamefont{Mart\'{\i}nez}},
  \bibinfo{journal}{Advances in Chemical Physics}
  \textbf{\bibinfo{volume}{121}}, \bibinfo{pages}{439} (\bibinfo{year}{2002}).

\bibitem[{\citenamefont{Heller}(2006)}]{heller2006guided}
\bibinfo{author}{\bibfnamefont{E.}~\bibnamefont{Heller}},
  \bibinfo{journal}{Acc. Chem. Res.} \textbf{\bibinfo{volume}{39}},
  \bibinfo{pages}{127} (\bibinfo{year}{2006}).

\bibitem[{\citenamefont{Worth et~al.}(2008)\citenamefont{Worth, Robb, and
  Lasorne}}]{worth2008solving}
\bibinfo{author}{\bibfnamefont{G.~A.} \bibnamefont{Worth}},
  \bibinfo{author}{\bibfnamefont{M.~A.} \bibnamefont{Robb}}, \bibnamefont{and}
  \bibinfo{author}{\bibfnamefont{B.}~\bibnamefont{Lasorne}},
  \bibinfo{journal}{Molecular Physics} \textbf{\bibinfo{volume}{106}},
  \bibinfo{pages}{2077} (\bibinfo{year}{2008}).

\bibitem[{\citenamefont{Richings et~al.}(2015)\citenamefont{Richings, Polyak,
  Spinlove, Worth, Burghardt, and Lasorne}}]{richings2015quantum}
\bibinfo{author}{\bibfnamefont{G.}~\bibnamefont{Richings}},
  \bibinfo{author}{\bibfnamefont{I.}~\bibnamefont{Polyak}},
  \bibinfo{author}{\bibfnamefont{K.}~\bibnamefont{Spinlove}},
  \bibinfo{author}{\bibfnamefont{G.}~\bibnamefont{Worth}},
  \bibinfo{author}{\bibfnamefont{I.}~\bibnamefont{Burghardt}},
  \bibnamefont{and} \bibinfo{author}{\bibfnamefont{B.}~\bibnamefont{Lasorne}},
  \bibinfo{journal}{International Reviews in Physical Chemistry}
  \textbf{\bibinfo{volume}{34}}, \bibinfo{pages}{269} (\bibinfo{year}{2015}).

\bibitem[{\citenamefont{Shalashilin}(2009)}]{shalashilin2009quantum}
\bibinfo{author}{\bibfnamefont{D.}~\bibnamefont{Shalashilin}},
  \bibinfo{journal}{J. Chem. Phys.} \textbf{\bibinfo{volume}{130}},
  \bibinfo{pages}{244101} (\bibinfo{year}{2009}).

\bibitem[{\citenamefont{Bonella and Coker}(2005)}]{Bonella2005}
\bibinfo{author}{\bibfnamefont{S.}~\bibnamefont{Bonella}} \bibnamefont{and}
  \bibinfo{author}{\bibfnamefont{D.~F.} \bibnamefont{Coker}},
  \bibinfo{journal}{J. Chem. Phys.} \textbf{\bibinfo{volume}{122}},
  \bibinfo{pages}{194102} (\bibinfo{year}{2005}).

\bibitem[{\citenamefont{Herman}(1984)}]{herman84}
\bibinfo{author}{\bibfnamefont{M.~F.} \bibnamefont{Herman}},
  \bibinfo{journal}{J. Chem. Phys.} \textbf{\bibinfo{volume}{81}},
  \bibinfo{pages}{754} (\bibinfo{year}{1984}).

\bibitem[{\citenamefont{Sun and Miller}(1997)}]{sun1997}
\bibinfo{author}{\bibfnamefont{X.}~\bibnamefont{Sun}} \bibnamefont{and}
  \bibinfo{author}{\bibfnamefont{W.~H.} \bibnamefont{Miller}},
  \bibinfo{journal}{J. Chem. Phys.} \textbf{\bibinfo{volume}{106}},
  \bibinfo{pages}{6346} (\bibinfo{year}{1997}).

\bibitem[{\citenamefont{Rassolov and Garashchuk}(2005)}]{PhysRevA.71.032511}
\bibinfo{author}{\bibfnamefont{V.~A.} \bibnamefont{Rassolov}} \bibnamefont{and}
  \bibinfo{author}{\bibfnamefont{S.}~\bibnamefont{Garashchuk}},
  \bibinfo{journal}{Phys. Rev. A} \textbf{\bibinfo{volume}{71}},
  \bibinfo{pages}{032511} (\bibinfo{year}{2005}).

\bibitem[{\citenamefont{Abedi et~al.}(2010)\citenamefont{Abedi, Maitra, and
  Gross}}]{Gross_PRL2010}
\bibinfo{author}{\bibfnamefont{A.}~\bibnamefont{Abedi}},
  \bibinfo{author}{\bibfnamefont{N.~T.} \bibnamefont{Maitra}},
  \bibnamefont{and} \bibinfo{author}{\bibfnamefont{E.~K.~U.}
  \bibnamefont{Gross}}, \bibinfo{journal}{Phys. Rev. Lett.}
  \textbf{\bibinfo{volume}{105}}, \bibinfo{pages}{123002}
  (\bibinfo{year}{2010}).

\bibitem[{\citenamefont{Abedi et~al.}(2012)\citenamefont{Abedi, Maitra, and
  Gross}}]{Gross_JCP2012}
\bibinfo{author}{\bibfnamefont{A.}~\bibnamefont{Abedi}},
  \bibinfo{author}{\bibfnamefont{N.~T.} \bibnamefont{Maitra}},
  \bibnamefont{and} \bibinfo{author}{\bibfnamefont{E.~K.~U.}
  \bibnamefont{Gross}}, \bibinfo{journal}{J. Chem. Phys.}
  \textbf{\bibinfo{volume}{137}}, \bibinfo{pages}{22A530}
  (\bibinfo{year}{2012}).

\bibitem[{\citenamefont{Abedi et~al.}(2013)\citenamefont{Abedi, Agostini,
  Suzuki, and Gross}}]{Gross_PRL2013}
\bibinfo{author}{\bibfnamefont{A.}~\bibnamefont{Abedi}},
  \bibinfo{author}{\bibfnamefont{F.}~\bibnamefont{Agostini}},
  \bibinfo{author}{\bibfnamefont{Y.}~\bibnamefont{Suzuki}}, \bibnamefont{and}
  \bibinfo{author}{\bibfnamefont{E.~K.~U.} \bibnamefont{Gross}},
  \bibinfo{journal}{Phys. Rev. Lett} \textbf{\bibinfo{volume}{110}},
  \bibinfo{pages}{263001} (\bibinfo{year}{2013}).

\bibitem[{\citenamefont{Agostini et~al.}(2013)\citenamefont{Agostini, Abedi,
  Suzuki, and Gross}}]{Gross_MP2013}
\bibinfo{author}{\bibfnamefont{F.}~\bibnamefont{Agostini}},
  \bibinfo{author}{\bibfnamefont{A.}~\bibnamefont{Abedi}},
  \bibinfo{author}{\bibfnamefont{Y.}~\bibnamefont{Suzuki}}, \bibnamefont{and}
  \bibinfo{author}{\bibfnamefont{E.~K.~U.} \bibnamefont{Gross}},
  \bibinfo{journal}{Mol. Phys.} \textbf{\bibinfo{volume}{111}},
  \bibinfo{pages}{3625} (\bibinfo{year}{2013}).

\bibitem[{\citenamefont{Agostini
  et~al.}(2015{\natexlab{a}})\citenamefont{Agostini, Abedi, Suzuki, Min,
  Maitra, and Gross}}]{Gross_JCP2015}
\bibinfo{author}{\bibfnamefont{F.}~\bibnamefont{Agostini}},
  \bibinfo{author}{\bibfnamefont{A.}~\bibnamefont{Abedi}},
  \bibinfo{author}{\bibfnamefont{Y.}~\bibnamefont{Suzuki}},
  \bibinfo{author}{\bibfnamefont{S.~K.} \bibnamefont{Min}},
  \bibinfo{author}{\bibfnamefont{N.~T.} \bibnamefont{Maitra}},
  \bibnamefont{and} \bibinfo{author}{\bibfnamefont{E.~K.~U.}
  \bibnamefont{Gross}}, \bibinfo{journal}{J. Chem. Phys.}
  \textbf{\bibinfo{volume}{142}}, \bibinfo{pages}{084303}
  (\bibinfo{year}{2015}{\natexlab{a}}).

\bibitem[{\citenamefont{Suzuki et~al.}(2015)\citenamefont{Suzuki, Abedi,
  Maitra, and Gross}}]{Suzuki_PCCP2015}
\bibinfo{author}{\bibfnamefont{Y.}~\bibnamefont{Suzuki}},
  \bibinfo{author}{\bibfnamefont{A.}~\bibnamefont{Abedi}},
  \bibinfo{author}{\bibfnamefont{N.~T.} \bibnamefont{Maitra}},
  \bibnamefont{and} \bibinfo{author}{\bibfnamefont{E.~K.~U.}
  \bibnamefont{Gross}}, \bibinfo{journal}{Phys. Chem. Chem. Phys.}
  \textbf{\bibinfo{volume}{17}}, \bibinfo{pages}{29271} (\bibinfo{year}{2015}).

\bibitem[{\citenamefont{Agostini
  et~al.}(2015{\natexlab{b}})\citenamefont{Agostini, Min, and
  Gross}}]{Agostini_ADP2015}
\bibinfo{author}{\bibfnamefont{F.}~\bibnamefont{Agostini}},
  \bibinfo{author}{\bibfnamefont{S.~K.} \bibnamefont{Min}}, \bibnamefont{and}
  \bibinfo{author}{\bibfnamefont{E.~K.~U.} \bibnamefont{Gross}},
  \bibinfo{journal}{Ann. Phys.} \textbf{\bibinfo{volume}{527}},
  \bibinfo{pages}{546} (\bibinfo{year}{2015}{\natexlab{b}}).

\bibitem[{\citenamefont{Scherrer et~al.}(2015)\citenamefont{Scherrer, Agostini,
  Sebastiani, Gross, and Vuilleumier}}]{Scherrer_JCP2015}
\bibinfo{author}{\bibfnamefont{A.}~\bibnamefont{Scherrer}},
  \bibinfo{author}{\bibfnamefont{F.}~\bibnamefont{Agostini}},
  \bibinfo{author}{\bibfnamefont{D.}~\bibnamefont{Sebastiani}},
  \bibinfo{author}{\bibfnamefont{E.~K.~U.} \bibnamefont{Gross}},
  \bibnamefont{and}
  \bibinfo{author}{\bibfnamefont{R.}~\bibnamefont{Vuilleumier}},
  \bibinfo{journal}{J. Chem. Phys.} \textbf{\bibinfo{volume}{143}},
  \bibinfo{pages}{074106} (\bibinfo{year}{2015}).

\bibitem[{\citenamefont{Schild et~al.}(2016)\citenamefont{Schild, Agostini, and
  Gross}}]{Schild_JPCA2016}
\bibinfo{author}{\bibfnamefont{A.}~\bibnamefont{Schild}},
  \bibinfo{author}{\bibfnamefont{F.}~\bibnamefont{Agostini}}, \bibnamefont{and}
  \bibinfo{author}{\bibfnamefont{E.~K.~U.} \bibnamefont{Gross}},
  \bibinfo{journal}{J. Phys. Chem. A} p.
  \bibinfo{pages}{10.1021/acs.jpca.5b12657} (\bibinfo{year}{2016}).

\bibitem[{\citenamefont{Eich and Agostini}(2016)}]{Agostini_arXiv2016}
\bibinfo{author}{\bibfnamefont{F.~G.} \bibnamefont{Eich}} \bibnamefont{and}
  \bibinfo{author}{\bibfnamefont{F.}~\bibnamefont{Agostini}},
  \bibinfo{journal}{arXiv:1604.05098 [physics.chem-ph]}
  (\bibinfo{year}{2016}).

\bibitem[{\citenamefont{Scherrer et~al.}(2016)\citenamefont{Scherrer, Agostini,
  Sebastiani, Gross, and Vuilleumier}}]{Scherrer_arXiv2016}
\bibinfo{author}{\bibfnamefont{A.}~\bibnamefont{Scherrer}},
  \bibinfo{author}{\bibfnamefont{F.}~\bibnamefont{Agostini}},
  \bibinfo{author}{\bibfnamefont{D.}~\bibnamefont{Sebastiani}},
  \bibinfo{author}{\bibfnamefont{E.~K.~U.} \bibnamefont{Gross}},
  \bibnamefont{and}
  \bibinfo{author}{\bibfnamefont{R.}~\bibnamefont{Vuilleumier}},
  \bibinfo{journal}{arXiv:1605.04211 [physics.chem-ph]}
  (\bibinfo{year}{2016}).

\bibitem[{\citenamefont{Hunter}(1974)}]{Hunter_IJQC1974}
\bibinfo{author}{\bibfnamefont{G.}~\bibnamefont{Hunter}},
  \bibinfo{journal}{Int. J. Quantum Chem.} \textbf{\bibinfo{volume}{8}},
  \bibinfo{pages}{413} (\bibinfo{year}{1974}).

\bibitem[{\citenamefont{Hunter}(1975{\natexlab{a}})}]{Hunter_IJQC1975_1}
\bibinfo{author}{\bibfnamefont{G.}~\bibnamefont{Hunter}},
  \bibinfo{journal}{Int. J. Quantum Chem.} \textbf{\bibinfo{volume}{9}},
  \bibinfo{pages}{237} (\bibinfo{year}{1975}{\natexlab{a}}).

\bibitem[{\citenamefont{Hunter}(1975{\natexlab{b}})}]{Hunter_IJQC1975_2}
\bibinfo{author}{\bibfnamefont{G.}~\bibnamefont{Hunter}},
  \bibinfo{journal}{Int. J. Quantum Chem.} \textbf{\bibinfo{volume}{9}},
  \bibinfo{pages}{311} (\bibinfo{year}{1975}{\natexlab{b}}).

\bibitem[{\citenamefont{Bishop and Hunter}(1975)}]{Hunter_MP1975}
\bibinfo{author}{\bibfnamefont{D.~M.} \bibnamefont{Bishop}} \bibnamefont{and}
  \bibinfo{author}{\bibfnamefont{G.}~\bibnamefont{Hunter}},
  \bibinfo{journal}{Mol. Phys.} \textbf{\bibinfo{volume}{30}},
  \bibinfo{pages}{1433} (\bibinfo{year}{1975}).

\bibitem[{\citenamefont{Bishop and Cheung}(1977)}]{Bishop_MP1975}
\bibinfo{author}{\bibfnamefont{D.~M.} \bibnamefont{Bishop}} \bibnamefont{and}
  \bibinfo{author}{\bibfnamefont{L.~M.} \bibnamefont{Cheung}},
  \bibinfo{journal}{Chem. Phys. Lett.} \textbf{\bibinfo{volume}{50}},
  \bibinfo{pages}{172} (\bibinfo{year}{1977}).

\bibitem[{\citenamefont{Hunter}(1980)}]{Hunter_IJQC1980}
\bibinfo{author}{\bibfnamefont{G.}~\bibnamefont{Hunter}},
  \bibinfo{journal}{Int. J. Quantum Chem.} \textbf{\bibinfo{volume}{9}},
  \bibinfo{pages}{133} (\bibinfo{year}{1980}).

\bibitem[{\citenamefont{Hunter}(1981)}]{Hunter_IJQC1981}
\bibinfo{author}{\bibfnamefont{G.}~\bibnamefont{Hunter}},
  \bibinfo{journal}{Int. J. Quantum Chem.} \textbf{\bibinfo{volume}{19}},
  \bibinfo{pages}{755} (\bibinfo{year}{1981}).

\bibitem[{\citenamefont{Hunter and Tai}(1982)}]{Hunter_IJQC1982}
\bibinfo{author}{\bibfnamefont{G.}~\bibnamefont{Hunter}} \bibnamefont{and}
  \bibinfo{author}{\bibfnamefont{C.~C.} \bibnamefont{Tai}},
  \bibinfo{journal}{Int. J. Quantum Chem.} \textbf{\bibinfo{volume}{21}},
  \bibinfo{pages}{1041} (\bibinfo{year}{1982}).

\bibitem[{\citenamefont{Hunter}(1986)}]{Hunter_IJQC1986}
\bibinfo{author}{\bibfnamefont{G.}~\bibnamefont{Hunter}},
  \bibinfo{journal}{Int. J. Quantum Chem.} \textbf{\bibinfo{volume}{29}},
  \bibinfo{pages}{197} (\bibinfo{year}{1986}).

\bibitem[{\citenamefont{Cederbaum}(2013)}]{Cederbaum_JCP2013}
\bibinfo{author}{\bibfnamefont{L.~S.} \bibnamefont{Cederbaum}},
  \bibinfo{journal}{J. Chem. Phys.} \textbf{\bibinfo{volume}{138}},
  \bibinfo{pages}{224110} (\bibinfo{year}{2013}).

\bibitem[{\citenamefont{Gidopoulos and Gross}(2014)}]{Gross_PTRSA2014}
\bibinfo{author}{\bibfnamefont{N.~I.} \bibnamefont{Gidopoulos}}
  \bibnamefont{and} \bibinfo{author}{\bibfnamefont{E.~K.~U.}
  \bibnamefont{Gross}}, \bibinfo{journal}{Phil. Trans. R. Soc. A}
  \textbf{\bibinfo{volume}{372}}, \bibinfo{pages}{20130059}
  (\bibinfo{year}{2014}).

\bibitem[{\citenamefont{Min et~al.}(2014)\citenamefont{Min, Abedi, Kim, and
  Gross}}]{Min_PRL2014}
\bibinfo{author}{\bibfnamefont{S.~K.} \bibnamefont{Min}},
  \bibinfo{author}{\bibfnamefont{A.}~\bibnamefont{Abedi}},
  \bibinfo{author}{\bibfnamefont{K.~S.} \bibnamefont{Kim}}, \bibnamefont{and}
  \bibinfo{author}{\bibfnamefont{E.~K.~U.} \bibnamefont{Gross}},
  \bibinfo{journal}{Phys. Rev. Lett.} \textbf{\bibinfo{volume}{113}},
  \bibinfo{pages}{263004} (\bibinfo{year}{2014}).

\bibitem[{\citenamefont{Chiang et~al.}(2014)\citenamefont{Chiang, Klaiman,
  Otto, and Cederbaum}}]{Cederbaum_JCP2014}
\bibinfo{author}{\bibfnamefont{Y.-C.} \bibnamefont{Chiang}},
  \bibinfo{author}{\bibfnamefont{S.}~\bibnamefont{Klaiman}},
  \bibinfo{author}{\bibfnamefont{F.}~\bibnamefont{Otto}}, \bibnamefont{and}
  \bibinfo{author}{\bibfnamefont{L.~S.} \bibnamefont{Cederbaum}},
  \bibinfo{journal}{J. Chem. Phys.} \textbf{\bibinfo{volume}{140}},
  \bibinfo{pages}{054104} (\bibinfo{year}{2014}).

\bibitem[{\citenamefont{Cederbaum}(2015)}]{Cederbaum_CP2015}
\bibinfo{author}{\bibfnamefont{L.~S.} \bibnamefont{Cederbaum}},
  \bibinfo{journal}{Chem. Phys.} \textbf{\bibinfo{volume}{457}},
  \bibinfo{pages}{129} (\bibinfo{year}{2015}).

\bibitem[{\citenamefont{Lefebvre}(2015{\natexlab{a}})}]{Lefebvre_JCP2015_1}
\bibinfo{author}{\bibfnamefont{R.}~\bibnamefont{Lefebvre}},
  \bibinfo{journal}{J. Chem. Phys.} \textbf{\bibinfo{volume}{142}},
  \bibinfo{pages}{074106} (\bibinfo{year}{2015}{\natexlab{a}}).

\bibitem[{\citenamefont{Lefebvre}(2015{\natexlab{b}})}]{Lefebvre_JCP2015_2}
\bibinfo{author}{\bibfnamefont{R.}~\bibnamefont{Lefebvre}},
  \bibinfo{journal}{J. Chem. Phys.} \textbf{\bibinfo{volume}{142}},
  \bibinfo{pages}{214105} (\bibinfo{year}{2015}{\natexlab{b}}).

\bibitem[{\citenamefont{Requist et~al.}(2015)\citenamefont{Requist, Tandetzky,
  and Gross}}]{Requist_PRA2015}
\bibinfo{author}{\bibfnamefont{R.}~\bibnamefont{Requist}},
  \bibinfo{author}{\bibfnamefont{F.}~\bibnamefont{Tandetzky}},
  \bibnamefont{and} \bibinfo{author}{\bibfnamefont{E.~K.~U.}
  \bibnamefont{Gross}}, \bibinfo{journal}{Phys. Rev. A} p.
  \bibinfo{pages}{Accepted} (\bibinfo{year}{2015}).

\bibitem[{\citenamefont{Abedi et~al.}(2014)\citenamefont{Abedi, Agostini, and
  Gross}}]{Gross_EPL2014}
\bibinfo{author}{\bibfnamefont{A.}~\bibnamefont{Abedi}},
  \bibinfo{author}{\bibfnamefont{F.}~\bibnamefont{Agostini}}, \bibnamefont{and}
  \bibinfo{author}{\bibfnamefont{E.~K.~U.} \bibnamefont{Gross}},
  \bibinfo{journal}{Europhys. Lett.} \textbf{\bibinfo{volume}{106}},
  \bibinfo{pages}{33001} (\bibinfo{year}{2014}).

\bibitem[{\citenamefont{Agostini et~al.}(2014)\citenamefont{Agostini, Abedi,
  and Gross}}]{Gross_JCP2014}
\bibinfo{author}{\bibfnamefont{F.}~\bibnamefont{Agostini}},
  \bibinfo{author}{\bibfnamefont{A.}~\bibnamefont{Abedi}}, \bibnamefont{and}
  \bibinfo{author}{\bibfnamefont{E.~K.~U.} \bibnamefont{Gross}},
  \bibinfo{journal}{J. Chem. Phys.} \textbf{\bibinfo{volume}{141}},
  \bibinfo{pages}{214101} (\bibinfo{year}{2014}).

\bibitem[{\citenamefont{Min et~al.}(2015)\citenamefont{Min, Agostini, and
  Gross}}]{Gross_PRL2015}
\bibinfo{author}{\bibfnamefont{S.~K.} \bibnamefont{Min}},
  \bibinfo{author}{\bibfnamefont{F.}~\bibnamefont{Agostini}}, \bibnamefont{and}
  \bibinfo{author}{\bibfnamefont{E.~K.~U.} \bibnamefont{Gross}},
  \bibinfo{journal}{Phys. Rev. Lett.} \textbf{\bibinfo{volume}{115}},
  \bibinfo{pages}{073001} (\bibinfo{year}{2015}).

\bibitem[{\citenamefont{Agostini et~al.}(2016)\citenamefont{Agostini, Min,
  Abedi, and Gross}}]{Gross_JCTC2016}
\bibinfo{author}{\bibfnamefont{F.}~\bibnamefont{Agostini}},
  \bibinfo{author}{\bibfnamefont{S.~K.} \bibnamefont{Min}},
  \bibinfo{author}{\bibfnamefont{A.}~\bibnamefont{Abedi}}, \bibnamefont{and}
  \bibinfo{author}{\bibfnamefont{E.~K.~U.} \bibnamefont{Gross}},
  \bibinfo{journal}{J. Chem. Theory Comput.} \textbf{\bibinfo{volume}{12}},
  \bibinfo{pages}{2127} (\bibinfo{year}{2016}).

\bibitem[{\citenamefont{Jeck et~al.}(2015)\citenamefont{Jeck, Sutcliffe, and
  Woolley}}]{Sutcliffe_TCA2012}
\bibinfo{author}{\bibfnamefont{T.}~\bibnamefont{Jeck}},
  \bibinfo{author}{\bibfnamefont{B.~T.} \bibnamefont{Sutcliffe}},
  \bibnamefont{and} \bibinfo{author}{\bibfnamefont{R.~G.}
  \bibnamefont{Woolley}}, \bibinfo{journal}{J. Phys. A: Math. Theor.}
  \textbf{\bibinfo{volume}{48}}, \bibinfo{pages}{445201}
  (\bibinfo{year}{2015}).

\bibitem[{\citenamefont{Suzuki et~al.}(2014)\citenamefont{Suzuki, Abedi,
  Maitra, Yamashita, and Gross}}]{Suzuki_PRA2014}
\bibinfo{author}{\bibfnamefont{Y.}~\bibnamefont{Suzuki}},
  \bibinfo{author}{\bibfnamefont{A.}~\bibnamefont{Abedi}},
  \bibinfo{author}{\bibfnamefont{N.~T.} \bibnamefont{Maitra}},
  \bibinfo{author}{\bibfnamefont{K.}~\bibnamefont{Yamashita}},
  \bibnamefont{and} \bibinfo{author}{\bibfnamefont{E.~K.~U.}
  \bibnamefont{Gross}}, \bibinfo{journal}{Phys. Rev. A}
  \textbf{\bibinfo{volume}{89}}, \bibinfo{pages}{040501(R)}
  (\bibinfo{year}{2014}).

\bibitem[{\citenamefont{Khosravi et~al.}(2015)\citenamefont{Khosravi, Abedi,
  and Maitra}}]{Khosravi_PRL2015}
\bibinfo{author}{\bibfnamefont{E.}~\bibnamefont{Khosravi}},
  \bibinfo{author}{\bibfnamefont{A.}~\bibnamefont{Abedi}}, \bibnamefont{and}
  \bibinfo{author}{\bibfnamefont{N.~T.} \bibnamefont{Maitra}},
  \bibinfo{journal}{Phys. Rev. Lett.} \textbf{\bibinfo{volume}{115}},
  \bibinfo{pages}{263002} (\bibinfo{year}{2015}).

\bibitem[{\citenamefont{Parashar et~al.}(2015)\citenamefont{Parashar, Sajeev,
  and Ghosh}}]{Ghosh_MP2015}
\bibinfo{author}{\bibfnamefont{S.}~\bibnamefont{Parashar}},
  \bibinfo{author}{\bibfnamefont{Y.}~\bibnamefont{Sajeev}}, \bibnamefont{and}
  \bibinfo{author}{\bibfnamefont{S.~K.} \bibnamefont{Ghosh}},
  \bibinfo{journal}{Mol. Phys.} \textbf{\bibinfo{volume}{113}},
  \bibinfo{pages}{3067} (\bibinfo{year}{2015}).

\bibitem[{\citenamefont{Granucci and Persico}(1995)}]{granucci1995coherent}
\bibinfo{author}{\bibfnamefont{G.}~\bibnamefont{Granucci}} \bibnamefont{and}
  \bibinfo{author}{\bibfnamefont{M.}~\bibnamefont{Persico}},
  \bibinfo{journal}{Chem. Phys. Lett.} \textbf{\bibinfo{volume}{246}},
  \bibinfo{pages}{228} (\bibinfo{year}{1995}).

\bibitem[{\citenamefont{Romstad et~al.}(1997)\citenamefont{Romstad, Granucci,
  and Persico}}]{romstad1997nonadiabatic}
\bibinfo{author}{\bibfnamefont{D.}~\bibnamefont{Romstad}},
  \bibinfo{author}{\bibfnamefont{G.}~\bibnamefont{Granucci}}, \bibnamefont{and}
  \bibinfo{author}{\bibfnamefont{M.}~\bibnamefont{Persico}},
  \bibinfo{journal}{Chem. Phys.} \textbf{\bibinfo{volume}{219}},
  \bibinfo{pages}{21} (\bibinfo{year}{1997}).

\bibitem[{\citenamefont{Donoso et~al.}(2000)\citenamefont{Donoso, Kohen, and
  Martens}}]{donoso2000simulation}
\bibinfo{author}{\bibfnamefont{A.}~\bibnamefont{Donoso}},
  \bibinfo{author}{\bibfnamefont{D.}~\bibnamefont{Kohen}}, \bibnamefont{and}
  \bibinfo{author}{\bibfnamefont{C.~C.} \bibnamefont{Martens}},
  \bibinfo{journal}{J. Chem. Phys.} \textbf{\bibinfo{volume}{112}},
  \bibinfo{pages}{7345} (\bibinfo{year}{2000}).

\bibitem[{\citenamefont{Granucci and Persico}(2007)}]{Granucci2007}
\bibinfo{author}{\bibfnamefont{G.}~\bibnamefont{Granucci}} \bibnamefont{and}
  \bibinfo{author}{\bibfnamefont{M.}~\bibnamefont{Persico}},
  \bibinfo{journal}{J. Chem. Phys.} \textbf{\bibinfo{volume}{126}},
  \bibinfo{pages}{134114} (\bibinfo{year}{2007}).

\bibitem[{\citenamefont{Granucci et~al.}(2010)\citenamefont{Granucci, Persico,
  and Zoccante}}]{granucci2010including}
\bibinfo{author}{\bibfnamefont{G.}~\bibnamefont{Granucci}},
  \bibinfo{author}{\bibfnamefont{M.}~\bibnamefont{Persico}}, \bibnamefont{and}
  \bibinfo{author}{\bibfnamefont{A.}~\bibnamefont{Zoccante}},
  \bibinfo{journal}{J. Chem. Phys.} \textbf{\bibinfo{volume}{133}},
  \bibinfo{pages}{134111} (\bibinfo{year}{2010}).

\bibitem[{\citenamefont{Shenvi et~al.}(2011{\natexlab{a}})\citenamefont{Shenvi,
  Subotnik, and Yang}}]{shenvi2011phase}
\bibinfo{author}{\bibfnamefont{N.}~\bibnamefont{Shenvi}},
  \bibinfo{author}{\bibfnamefont{J.}~\bibnamefont{Subotnik}}, \bibnamefont{and}
  \bibinfo{author}{\bibfnamefont{W.}~\bibnamefont{Yang}}, \bibinfo{journal}{J.
  Chem. Phys.} \textbf{\bibinfo{volume}{135}}, \bibinfo{pages}{024101}
  (\bibinfo{year}{2011}{\natexlab{a}}).

\bibitem[{\citenamefont{Subotnik and
  Shenvi}(2011{\natexlab{a}})}]{subotnik2011decoherence}
\bibinfo{author}{\bibfnamefont{J.}~\bibnamefont{Subotnik}} \bibnamefont{and}
  \bibinfo{author}{\bibfnamefont{N.}~\bibnamefont{Shenvi}},
  \bibinfo{journal}{J. Chem. Phys.} \textbf{\bibinfo{volume}{134}},
  \bibinfo{pages}{244114} (\bibinfo{year}{2011}{\natexlab{a}}).

\bibitem[{\citenamefont{Shenvi et~al.}(2011{\natexlab{b}})\citenamefont{Shenvi,
  Subotnik, and Yang}}]{shenvi2011simultaneous}
\bibinfo{author}{\bibfnamefont{N.}~\bibnamefont{Shenvi}},
  \bibinfo{author}{\bibfnamefont{J.}~\bibnamefont{Subotnik}}, \bibnamefont{and}
  \bibinfo{author}{\bibfnamefont{W.}~\bibnamefont{Yang}}, \bibinfo{journal}{J.
  Chem. Phys.} \textbf{\bibinfo{volume}{134}}, \bibinfo{pages}{144102}
  (\bibinfo{year}{2011}{\natexlab{b}}).

\bibitem[{\citenamefont{Subotnik and
  Shenvi}(2011{\natexlab{b}})}]{subotnik2011new}
\bibinfo{author}{\bibfnamefont{J.}~\bibnamefont{Subotnik}} \bibnamefont{and}
  \bibinfo{author}{\bibfnamefont{N.}~\bibnamefont{Shenvi}},
  \bibinfo{journal}{J. Chem. Phys.} \textbf{\bibinfo{volume}{134}},
  \bibinfo{pages}{024105} (\bibinfo{year}{2011}{\natexlab{b}}).

\bibitem[{\citenamefont{Shenvi and Yang}(2012)}]{shenvi2012achieving}
\bibinfo{author}{\bibfnamefont{N.}~\bibnamefont{Shenvi}} \bibnamefont{and}
  \bibinfo{author}{\bibfnamefont{W.}~\bibnamefont{Yang}}, \bibinfo{journal}{J.
  Chem. Phys.} \textbf{\bibinfo{volume}{137}}, \bibinfo{pages}{22A528}
  (\bibinfo{year}{2012}).

\bibitem[{\citenamefont{Born and Huang}(1954)}]{bornhuang}
\bibinfo{author}{\bibfnamefont{M.}~\bibnamefont{Born}} \bibnamefont{and}
  \bibinfo{author}{\bibfnamefont{K.}~\bibnamefont{Huang}},
  \emph{\bibinfo{title}{Dynamical Theory of Crystal Lattices}}
  (\bibinfo{publisher}{Clarendon, Oxford}, \bibinfo{year}{1954}).

\bibitem[{\citenamefont{Barbatti et~al.}(2011)\citenamefont{Barbatti, Shepard,
  and Lischka}}]{barbatti10a}
\bibinfo{author}{\bibfnamefont{M.}~\bibnamefont{Barbatti}},
  \bibinfo{author}{\bibfnamefont{R.}~\bibnamefont{Shepard}}, \bibnamefont{and}
  \bibinfo{author}{\bibfnamefont{H.}~\bibnamefont{Lischka}}, in
  \emph{\bibinfo{booktitle}{Conical Intersections: Theory, Computation and
  Experiment}}, edited by
  \bibinfo{editor}{\bibfnamefont{W.}~\bibnamefont{Domcke}},
  \bibinfo{editor}{\bibfnamefont{D.~R.} \bibnamefont{Yarkony}},
  \bibnamefont{and} \bibinfo{editor}{\bibfnamefont{H.}~\bibnamefont{Koeppel}}
  (\bibinfo{publisher}{Singapore, World Scientific}, \bibinfo{year}{2011}), p.
  \bibinfo{pages}{415}.

\bibitem[{\citenamefont{Curchod et~al.}(2013)\citenamefont{Curchod,
  Rothlisberger, and Tavernelli}}]{curchod2013trajectory}
\bibinfo{author}{\bibfnamefont{B.~F.~E.} \bibnamefont{Curchod}},
  \bibinfo{author}{\bibfnamefont{U.}~\bibnamefont{Rothlisberger}},
  \bibnamefont{and}
  \bibinfo{author}{\bibfnamefont{I.}~\bibnamefont{Tavernelli}},
  \bibinfo{journal}{ChemPhysChem} \textbf{\bibinfo{volume}{14}},
  \bibinfo{pages}{1314} (\bibinfo{year}{2013}).

\bibitem[{\citenamefont{Malhado et~al.}(2014)\citenamefont{Malhado, Bearpark,
  and Hynes}}]{malhado2014non}
\bibinfo{author}{\bibfnamefont{J.~P.} \bibnamefont{Malhado}},
  \bibinfo{author}{\bibfnamefont{M.~J.} \bibnamefont{Bearpark}},
  \bibnamefont{and} \bibinfo{author}{\bibfnamefont{J.~T.} \bibnamefont{Hynes}},
  \bibinfo{journal}{Frontiers in chemistry} \textbf{\bibinfo{volume}{2}},
  \bibinfo{pages}{97} (\bibinfo{year}{2014}).

\bibitem[{\citenamefont{de~Carvalho et~al.}(2014)\citenamefont{de~Carvalho,
  Bouduban, Curchod, and Tavernelli}}]{e16010062}
\bibinfo{author}{\bibfnamefont{F.~F.} \bibnamefont{de~Carvalho}},
  \bibinfo{author}{\bibfnamefont{M.~E.~F.} \bibnamefont{Bouduban}},
  \bibinfo{author}{\bibfnamefont{B.~F.~E.} \bibnamefont{Curchod}},
  \bibnamefont{and}
  \bibinfo{author}{\bibfnamefont{I.}~\bibnamefont{Tavernelli}},
  \bibinfo{journal}{Entropy} \textbf{\bibinfo{volume}{16}}, \bibinfo{pages}{62}
  (\bibinfo{year}{2014}), ISSN \bibinfo{issn}{1099-4300}.

\bibitem[{\citenamefont{Shevchenko et~al.}(2010)\citenamefont{Shevchenko,
  Ashhab, and Nori}}]{shevchenko2010landau}
\bibinfo{author}{\bibfnamefont{S.}~\bibnamefont{Shevchenko}},
  \bibinfo{author}{\bibfnamefont{S.}~\bibnamefont{Ashhab}}, \bibnamefont{and}
  \bibinfo{author}{\bibfnamefont{F.}~\bibnamefont{Nori}},
  \bibinfo{journal}{Phys. Rep.} \textbf{\bibinfo{volume}{492}},
  \bibinfo{pages}{1} (\bibinfo{year}{2010}).

\bibitem[{\citenamefont{Shin and Metiu}(1995)}]{shin1995nonadiabatic}
\bibinfo{author}{\bibfnamefont{S.}~\bibnamefont{Shin}} \bibnamefont{and}
  \bibinfo{author}{\bibfnamefont{H.}~\bibnamefont{Metiu}}, \bibinfo{journal}{J.
  Chem. Phys.} \textbf{\bibinfo{volume}{102}}, \bibinfo{pages}{9285}
  (\bibinfo{year}{1995}).

\bibitem[{\citenamefont{Feit et~al.}(1982)\citenamefont{Feit, {Fleck Jr.}, and
  Steiger}}]{spo}
\bibinfo{author}{\bibfnamefont{M.~D.} \bibnamefont{Feit}},
  \bibinfo{author}{\bibfnamefont{F.~A.} \bibnamefont{{Fleck Jr.}}},
  \bibnamefont{and} \bibinfo{author}{\bibfnamefont{A.}~\bibnamefont{Steiger}},
  \bibinfo{journal}{J. Comput. Phys.} \textbf{\bibinfo{volume}{47}},
  \bibinfo{pages}{412} (\bibinfo{year}{1982}).

\bibitem[{\citenamefont{Deumens et~al.}(1994)\citenamefont{Deumens, Diz, Longo,
  and {\"O}hrn}}]{deumens1994time}
\bibinfo{author}{\bibfnamefont{E.}~\bibnamefont{Deumens}},
  \bibinfo{author}{\bibfnamefont{A.}~\bibnamefont{Diz}},
  \bibinfo{author}{\bibfnamefont{R.}~\bibnamefont{Longo}}, \bibnamefont{and}
  \bibinfo{author}{\bibfnamefont{Y.}~\bibnamefont{{\"O}hrn}},
  \bibinfo{journal}{Rev. Mod. Phys.} \textbf{\bibinfo{volume}{66}},
  \bibinfo{pages}{917} (\bibinfo{year}{1994}).

\bibitem[{\citenamefont{Lopreore and Wyatt}(2002)}]{Lopreore2002}
\bibinfo{author}{\bibfnamefont{C.~L.} \bibnamefont{Lopreore}} \bibnamefont{and}
  \bibinfo{author}{\bibfnamefont{R.~E.} \bibnamefont{Wyatt}},
  \bibinfo{journal}{J. Chem. Phys.} \textbf{\bibinfo{volume}{116}},
  \bibinfo{pages}{1228} (\bibinfo{year}{2002}).

\bibitem[{\citenamefont{Takatsuka}(2006)}]{takatsuka2006non}
\bibinfo{author}{\bibfnamefont{K.}~\bibnamefont{Takatsuka}},
  \bibinfo{journal}{J. Chem. Phys.} \textbf{\bibinfo{volume}{124}},
  \bibinfo{pages}{064111} (\bibinfo{year}{2006}).

\bibitem[{\citenamefont{Micha and Burghardt}(2007)}]{irenebook}
\bibinfo{editor}{\bibfnamefont{D.~A.} \bibnamefont{Micha}} \bibnamefont{and}
  \bibinfo{editor}{\bibfnamefont{I.}~\bibnamefont{Burghardt}}, eds.,
  \emph{\bibinfo{title}{Quantum Dynamics of Complex Systems}}
  (\bibinfo{publisher}{Springer Series in Chemical Physics},
  \bibinfo{year}{2007}).

\bibitem[{\citenamefont{Shenvi}(2009)}]{shenvi2009phase}
\bibinfo{author}{\bibfnamefont{N.}~\bibnamefont{Shenvi}}, \bibinfo{journal}{J.
  Chem. Phys.} \textbf{\bibinfo{volume}{130}}, \bibinfo{pages}{124117}
  (\bibinfo{year}{2009}).

\end{thebibliography}

\end{document}